\documentclass[12pt]{article}
\usepackage{amsfonts, amsmath, amssymb, cite, graphicx}

\usepackage{color}
\usepackage[all, knot]{xy}
\usepackage{tikz}
\usepackage{braket}

\usepackage[utf8]{inputenc}

\usepackage{epstopdf}
\usepackage[footnotesize]{caption}
\usepackage{amsthm}
\usepackage{enumitem}
\usepackage{mathrsfs}

\usepackage[margin=3cm]{geometry}

\def \be {\begin{equation}}
\def \ee {\end{equation}}
\def \ba {\begin{aligned}}
\def \ea {\end{aligned}}
\def \bea {\begin{eqnarray}}
\def \eea {\end{eqnarray}}

\def\CI{{\cal I}}
\def\CJ{{\cal J}}

\def\CN{{\cal N}}
\def\CO{{\cal O}}

\def\>{\rangle} 
\def\<{\langle} 
\def\vev#1{\langle #1\rangle}

\def\s{\sigma}

\def\th{\theta}
\def\tt{\tilde{\theta}}
\def\btt{\bar{\tilde{\theta}}}
\def\l{\lambda}

\def\p{\partial}

\def\a{\alpha}

\begin{document}
\begin{titlepage}
\vspace{0.5cm}
\begin{center}
{\Large \bf The correlation function of (1,1) and (2,2) supersymmetric theories with $T\bar{T}$ deformation  }

\lineskip .75em
\vskip 2.5cm
{\large Song He$^{a,b,}$\footnote{hesong@jlu.edu.cn}, Jia-Rui Sun$^{c,d,}$\footnote{sunjiarui@sysu.edu.cn}, Yuan Sun$^{c,}$\footnote{sunyuan6@mail.sysu.edu.cn}}
\vskip 2.5em
 {\normalsize\it $^{a}$Center for Theoretical Physics and College of Physics, Jilin University, Changchun 130012, People's Republic of China\\
 $^{b}$Max Planck Institute for Gravitational Physics (Albert Einstein Institute),\\
Am M\"uhlenberg 1, 14476 Golm, Germany\\
$^{c}$   School of Physics and Astronomy, Sun Yat-Sen University, Guangzhou 510275, China\\
$^{d}$   Guangdong Provincial Key Laboratory of Quantum Metrology and Sensing, Sun Yat-Sen University, Guangzhou 510275, China}\\
\vskip 3.0em
\end{center}
\begin{abstract}
In the paper, we compute the correlation functions in 2D $\CN=(1,1)$ and $\CN=(2,2)$  superconformal field theories with $T\bar{T}$ deformation up to the first order of the deformation in terms of perturbation theory. With the help of superconformal Ward identity in $\CN=(1,1)$ and $\CN=(2,2)$ theories and careful regularization, the correlation functions in the deformed theory can be obtained up to the first order perturbation.  This study is the extension from previous bosonic $T\bar{T}$ deformation to the supersymmetric one.
  \end{abstract}
\end{titlepage}

\baselineskip=0.7cm

\tableofcontents
\newpage

\section{Introduction}
Studying exactly solvable models in 2D QFT can help us get a deep understanding of general field theory. The next natural step is to consider the deviation from these exactly solvable models. In the language of renormalization group (RG) flow, when turning on deformations, usually the relevant ones are more controllable than the irrelevant ones since the latter may introduce infinite divergences in the UV. However, a special irrelevant deformation
has shown a number of remarkable properties even in the UV \cite{Zamolodchikov:2004ce,Smirnov:2016lqw,Cavaglia:2016oda}. Such deformation preserves the integrability if the undeformed theory is integrable, which makes it convenient to obtain the spectrum and S-matrix. In addition, the deformed theory was shown can be renormalized in a systematical way \cite{Rosenhaus:2019utc}.

Recently, the $T\bar{T}$ deformation as a special deformation, has attracted much attention \cite{Aharony:2018vux,Datta:2018thy,Aharony:2018bad,Conti:2018tca,Bonelli:2018kik,Dubovsky:2018bmo,Conti:2018jho,Santilli:2018xux,Jiang:2019tcq,Conti:2019dxg,Cardy:2018jho,Guica:2017lia,Guica:2019vnb,Cardy:2019qao,He:2019vzf,Donnelly:2018bef,Wang:2018jva,Sun:2019ijq,Jiang:2019hxb,Grieninger:2019zts,Jeong:2019ylz,Banerjee:2019ewu,Gross:2019ach,Gross:2019uxi}. Here $T$ is stress tensor of the theory. The deformed Lagrangian $S(\l)$ can be written as
\be \label{TTLa}
\frac{\p S(\l)}{\p \l}=\int d^2z T\bar{T}(z),
\ee
where the operator $T\bar{T}(z)$ was first introduced in \cite{Zamolodchikov:2004ce}. For conformal field theory ( CFT) on torus, the partition function with the deformation can be computed and it remains modular invariant. Further, one can even obtain Cardy-like formula in deformed CFTs. Meanwhile, there  were also various other perspectives on the $T\bar{T}$ deformation\cite{Dubovsky:2017cnj,Cardy:2018sdv,Giveon:2017nie,Giveon:2017myj,Asrat:2017tzd,Giribet:2017imm,Baggio:2018gct,Apolo:2018qpq,Babaro:2018cmq,Chakraborty:2018aji,Araujo:2018rho,Giveon:2019fgr,Chakraborty:2019mdf,Apolo:2019zai}. In particular, from a holographic point of view, it was suggested that with coupling constant $\l>0$ in eq.(\ref{TTLa}), the $T\bar{T}$ deformed 2D CFT is dual to an AdS$_3$ gravity with finite cutoff in the radial direction \cite{McGough:2016lol,Guica:2019nzm}. Evidences for this dictionary are associated with matching of the energy spectrum, holographic entanglement entropies, exact holographic renormalization and so on. For recent progresses on holographic aspects  of $T\bar{T}$ deformation, one can also refer to \cite{Shyam:2017znq,Kraus:2018xrn,Cottrell:2018skz,Bzowski:2018pcy,Taylor:2018xcy,Hartman:2018tkw,Shyam:2018sro,Caputa:2019pam,Apolo:2019yfj,Chen:2018eqk,Chen:2019mis,He:2019glx}.

There are many directions to generalize the $T\bar{T}$ deformation. An interesting question is that what will happen when additional symmetry is presented in the theory, for example, the supersymmetry (SUSY). In \cite{Baggio:2018rpv,Chang:2018dge,Jiang:2019hux,Chang:2019kiu} (see also \cite{Coleman:2019dvf,Jiang:2019trm}), the authors have taken the SUSY into account, for examples, $\mathcal{N}=(0,1)$ and extend SUSY with $\CN=(1,1),(2,0),(2,2)$ . In these studies, the supersymmetric extensions of $T\bar{T}$ operator presented in eq.(\ref{TTLa}) was constructed based on the supercurrent multiplet \cite{Dumitrescu:2011iu}, and the deformed Lagrangian was also given for free theory with or without the potential. Taking $\CN=(1,1)$ for example \cite{Baggio:2018rpv}, the deformed action takes the form
\be \label{susyTL}
S_\a=S_0+\l\int d^2\s O(\s)
\ee
with
\be
O(\s)=\int d\theta^+ d\theta^- \CO(\zeta),
\ee
where $\CO(\xi)=\CJ_{+++}(\zeta)\CJ_{---}(\zeta)-\CJ_-(\zeta)\CJ_+(\zeta)$, $(\CJ_{+++},\CJ_-)$ and $(\CJ_{---},\CJ_+)$ are two pairs of superfields, which include the stress energy tensor (For more details for this construction, one can refer to \cite{Baggio:2018rpv}). Moreover, it was shown that the deformation constructed in this way preserves  both the solvability and  the SUSY. Furthermore, the $O$ in eq.(\ref{susyTL}) is a composite operator which is equivalent to the $T\bar{T}$ as presented in eq.(\ref{TTLa}) at the classical level up to some total derivative terms
\be
O =T\bar{T} +\text{EOM}'s+\text{total derivatives}.
\ee

In the present paper, we are interested in studying the correlation functions of the $T\bar{T}$ deformed superconformal field theory, perturbatively. The correlation functions have been perturbatively investigated in $T\bar{T}$ \cite{Kraus:2018xrn,He:2019vzf} and $J\bar{T}$ \cite{Guica:2019vnb} deformed theories without SUSY, and they were also proposed in a non-perturbative way by J. Cardy \cite{Cardy:2019qao}. Since we will work in the Euclidean signature, we will focus on the correlation functions of the deformed superconformal field theory with $\CN=(1,1)$ and $\CN=(2,2)$ SUSY. Due to the on-shell condition, the operators $O$ and $T\bar{T}$ are equivalent up to some total derivative terms at the classical level. For convenience, we will directly employ $T\bar{T}$ instead of $O$ in the processes of computing the correlation functions. Here we have to emphasize that we only focus on small deformations away from the undeformed CFTs, such that the CFT Ward identity still holds and it is not necessary to take account the effect of the RG flow of the operator with the irrelevant deformation. Therefore, the conformal symmetry can be considered as an approximate symmetry up to the first order of the $T\bar{T}$ deformation and the correlation functions can also be  obtained perturbatively. Moreover, both in holography and quantum field theory, these correlation functions can also be applied to obtain various interesting quantum information quantities in the deformed field theory, such as the R\'enyi entropy of local quench in various situations \cite{Nozaki:2014hna,He:2014mwa,He:2017lrg}, entanglement negativity \cite{Calabrese:2012ew}, entanglement purification \cite{Takayanagi:2017knl}, information metric\cite{MIyaji:2015mia,Suzuki:2019xdq}, etc.

The remaining parts of the paper are organized as follows. In section 2, we first briefly review the Ward identity in (1,1) superconformal symmetry and also the correlation functions in undeformed theory, then formulate the 2-,3-, and $n$-point ($n$-pt) correlation functions with $T\bar{T}$ inserted, the last step is to perform the integral in conformal perturbation theory using dimensional regularization. In section 3, we first discuss the Ward identity and undeformed correlators in (2,2) superconformal field theory.  Then following the same line as section 2, we compute the  2-,3-, and $n$-point deformed correlation functions. In section 4. We discuss the dimensional regularization methods used in section 2 and section 3. In the final section, conclusions and discussions will be given. In appendices, we would like to list some techniques and relevant notations which are very useful in our analysis.

\section{$\CN$=(1,1) superconformal symmetry}
In this section we review (1,1) superconformal symmetry and the corresponding Ward identity. The coordinates on superspace are analytic coordinates $Z=(z,\theta)$ and anti-analytic coordinates $\bar{Z}=(\bar{z},\bar{\theta})$ where $z,\bar{z}$ are two complex coordinates and $\theta,\bar{\theta}$ are Grassmannian coordinates. The (1,1) superconformal algebra is the direct sum of the (1,0) and the (0,1) algebras, thus for simplicity, we will only write out its analytic part in the following. For (1,1) theory the superderivative is \cite{West:1990tg,Blumenhagen:2009zz,Fuchs:1986ew,Petersen:1985bs}
\be
D=\p_\theta+\theta \p_z,~~D^2=\p_z.
\ee
The superfield
\be
J(Z)=\Theta(z)+\theta T(z)
\ee
generates analytic supercoordinates transformations in superspace. Here $T(z)$  is stress-energy tensor of the theory and $\Theta$ is a generator of SUSY transformations. Similar expression can be write out for $\bar{J}(Z)$.

Under analytic supercoordinates transformations with parameter $E(Z)$, a local superfield  $\Phi(Z,\bar{Z})$ obeys
\be \label{delp}
\delta_E \Phi(Z,\bar{Z})=[J_E,\Phi(Z,\bar{Z})]=\oint dZ'E(Z')J(Z')\Phi(Z,\bar{Z})
\ee
with
\be
\oint dZ\equiv \frac{1}{2\pi i}\oint dz\int d\theta.
\ee

A superfield $\Phi(Z,\bar{Z})$ is called primary superfield if it transforms as
\be \label{varia}
\delta_E\Phi(Z,\bar{Z})=E(Z)\p_z\Phi(Z,\bar{Z})+\frac{1}{2}DE(Z)D\Phi(Z,\bar{Z})+\Delta \p_z E(Z)\Phi(Z,\bar{Z}),
\ee
where only the analytic part of the transformation is considered. Furthermore, one can obtain the OPE  between the superfield $J(Z)$ containing stress tensor $T(z)$ and primary superfield $\Phi$ with dimension $\Delta$, which is the generalization of OPE between stress tensor and primary field $T(z)\phi(z')$ in CFT.
This can be done
by substituting eq.(\ref{varia}) back to eq.(\ref{delp}) and  using the identities
\be
\oint dZ_1 E(Z_1)\frac{\theta_{12}}{Z_{12} } =E(Z_2),
\ee
\be
\oint dZ_1 E(Z_1)\frac{1}{Z_{12} } =DE(Z_2),
\ee
\be
\oint dZ_1 E(Z_1)\frac{\theta_{12}}{Z_{12}^2} =\p_z E(Z_2),
\ee
where the SUSY invariant distance $Z_{12}=z_1-z_2-\theta_1\theta_2$ and $\theta_{12}=\theta_1-\theta_2$. Note these identities can be obtained by super-Cauchy theorem
 \be
\oint dZ'E(Z')\frac{\theta'-\theta}{Z'-Z}=E(Z).
\ee
We then obtain the following OPE \cite{Fuchs:1986ew}
\be
J(Z_1)\Phi(Z_2)=\frac{\theta_{12}}{Z_{12}}\p_{z_2}\Phi(Z_2,\bar{Z}_2)+\frac{1}{2}\frac{1}{Z_{12}}D\Phi(Z_2,\bar{Z}_2)+\Delta\frac{\theta_{12}}{Z^2_{12}}\Phi(Z_2,\bar{Z}_2).
\ee
From this OPE, the $\CN=(1,1)$ superconformal Ward identity can be written as
\be\ba
&\vev{J(Z_0)\Phi_1(Z_1,\bar{Z}_1)...\Phi_n(Z_n,\bar{Z}_n)}\\
=&\sum_{i=1}^n\Big(\frac{\theta_{0i}}{Z_{0i} }\p_{z_i}+\frac{1}{2Z_{0i} } D_i+ \Delta_i \frac{\theta_{0i}}{Z_{0i} ^2} \Big)\vev{ \Phi_1(Z_1,\bar{Z}_1)...\Phi_n(Z_n,\bar{Z}_n)}.
\ea\ee
and similar expressions for $\bar{J}(\bar{Z})$.

It is important to apply the Ward identity to global superconformal transformation whose algebra osp(2$|$1) is a subalgebra of superconformal algebra. By employing the Ward identity and the fact that correlator of primary superfields is invariant under global  superconformal transformation since it is a true symmetry of the theory, these correlators will be highly constrained. And similar to the cases in bosonic CFT, it is possible to completely fix 2- and  3-point correlators up to some constant factors. The 2-pt correlator is
\be \label{2pt}
\vev{\Phi_1(Z_1,\bar{Z}_1)\Phi_2(Z_2,\bar{Z}_2)}=c_{12}\frac{1}{Z_{12}^{2\Delta}\bar{Z}_{12}^{2\bar{\Delta}}},~~\Delta\equiv\Delta_1=\Delta_2,~~\bar{\Delta}\equiv\bar{\Delta}_1=\bar{\Delta}_2
\ee
with $c_{12}$ a constant  and 3-pt correlator is
\be \label{3pt}
\vev{\Phi_1(Z_1,\bar{Z}_1)\Phi_2(Z_2,\bar{Z}_2)\Phi_3(Z_3,\bar{Z}_3)}=\Big(\prod_{i<j=1}^3 \frac{1}{Z_{ij}^{\Delta_{ij}}\bar{Z}_{ij}^{\bar{\Delta}_{ij}}}\Big)(c_{123}+c'_{123}\theta_{123}\bar{\theta}_{123}),
\ee
where the second factor in the right hand side can also be written as
\be \label{et}
c_{123}+c'_{123}\theta_{123}\bar{\theta}_{123}=c_{123}e^{c'_{123}\theta_{123}\bar{\theta}_{123}/c_{123}}.
\ee
Here $c_{123},c'_{123}$ are constants, $\Delta_{ij}=\Delta_i+\Delta_j-\epsilon_{ijk}\Delta_k,(i<j)$ (for $i>j$, we define $\Delta_{ij}\equiv \Delta_{ji}$ such that $\Delta_{ij}$ is symmetric), and $\theta_{123}$ is defined as
\be \label{thetaijk}
\theta_{ijk}=\frac{1}{\sqrt{Z_{ij}Z_{jk}Z_{kl}}}(\theta_i Z_{jk}+\theta_j Z_{ki}+\theta_k Z_{ij}+\theta_i\theta_j\theta_k),
\ee
which is invariant under global conformal  transformation. By definition $\theta_{123}$ is Grassmann-odd, thus $\theta_{123}^2=0$ and eq.(\ref{et}) holds.

As for $n$-pt correlators with $n\geq 4$, they depend on $2n$ coordinates $z_i,\theta_i,i=1,...,n$, and 5 constraints corresponding to 5 generators of osp(2$|$1). Thus there are $2n-5$ independent variables in $n$-pt correlators. Actually, there exists the same number of independent osp(2$|$1) invariants, i.e. $2n-5$, which are \cite{Fuchs:1986ew}
\be
w_j\equiv\theta_{12j},~j=3,...,n,~~~U_k\equiv Z_{123k},~k=4,...,n,
\ee
where $\theta_{12j}$ is defined in eq.(\ref{thetaijk}) and $Z_{ijkl}$ is an analogue of cross ratio in CFT
\be
Z_{ijkl}=\frac{Z_{ij}Z_{kl}}{Z_{li}Z_{jk}}.
\ee
The $n$-pt function can be determined in terms of these variables as
\be \label{npt}
\vev{\Phi_1(Z_1,\bar{Z}_1)...\Phi_n(Z_n,\bar{Z}_n)}=\Big(\prod_{i<j=1}^n \frac{1}{Z_{ij}^{\Delta_{ij}}\bar{Z}_{ij}^{\bar{\Delta}_{ij}}}\Big)f(w_i,\bar{w}_i,U_j,\bar{U}_j)
\ee
with $\sum_{i\neq j}\Delta_{ij}=2\Delta_j,\Delta_{ij}=\Delta_{ji}$ and similar for $\bar{\Delta}_{ij}$. Here $f$ is a function which can not be fixed by global superconformal symmetry, and it depends on the theory under consideration.

With the results discussed above, we can compute the $T\bar{T}$ deformed correlators. The variation of action under $T\bar{T}$ deformation can be constructed as
\be
\delta S=\l\int d^2zT\bar{T}(z)=-\l\int d^2z\int d\theta d\bar{\theta}J(Z)\bar{J}(\bar{Z}),
\ee
where the minus sign comes from the anti-commutation nature of $\theta$. Thus up to first order in $\l$ the variation of $n$-pt correlator  is
\be \label{TTbar}
-\l\int d^2z\int d\theta d\bar{\theta}\vev{J(Z)\bar{J}(\bar{Z})\Phi(Z_1,\bar{Z}_1)...\Phi(Z_n,\bar{Z}_n)}.
\ee
Note that the correlator inside the integral can be evaluated via the Ward identity. In the following  subsections we will compute  eq.(\ref{TTbar}) for $n=2,3$ and $n\geq 4$.

\subsection{2-point correlators}
In this subsection we will consider the 2-point correlators with $T\bar{T}$ deformation. The undeformed correlator takes the form as eq.(\ref{2pt})
\be \label{2pt1}
\vev{\Phi_1(Z_1,\bar{Z}_1)\Phi_2(Z_2,\bar{Z}_2)}=\frac{c_{12}}{Z_{12}^{2\Delta}\bar{Z}_{12}^{2\bar{\Delta}}}.
\ee
First, considering the case with only holomorphic component of stress tensor inserted in above correlator
\be\ba
&\vev{J(Z)\Phi_1\Phi_2}= \sum_{i=1}^2\Big(\frac{\theta_{0i}}{Z_{0i}}\p_{z_i}+\frac{1}{2Z_{0i}}D_i+\frac{\Delta_i\theta_{0i}}{Z^2_{0i}} \Big)\vev{\Phi_1\Phi_2},
\ea\ee
where $\th_{0i}=\theta-\theta_i,Z_{0i}=z-z_i-\theta\theta_i$, and  the derivatives on the right hand side can act on both holomorphic and antiholomorphic parts of $\vev{\Phi_1\Phi_2}$. For example, for holomorphic part
\be\ba
\p_{z_1}\frac{1}{Z_{12}^{2\Delta} }=-2\Delta\frac{1}{Z_{12}^{2\Delta+1}},
~~
D_1\frac{1}{Z_{12}^{2\Delta} }=-2\Delta\frac{\theta_{12}}{Z_{12}^{2\Delta+1}},
\ea\ee
and for antiholomorphic part\footnote{Useful formulae
\be \label{expand}
\frac{1}{\bar{Z}_{ij}}=\frac{1}{\bar{z}_{ij}}+\frac{\bar{\theta}_1\bar{\theta}_2}{\bar{z}^2_{ij}},~~\frac{\theta_{ij}}{Z_{ij}}=\frac{\theta_{ij}}{z_{ij}},~~\frac{\theta_{i}}{Z_{ij}}=\frac{\theta_{i}}{z_{ij}}.
\ee
And the differential
\be
\p_{z_1}\frac{1}{\bar{Z}_{12}}=\p_{z_1}\Big(\frac{1}{\bar{z}_{12}}+\frac{\bar{\theta}_1\bar{\theta}_2}{\bar{z}^2_{12}}\Big) =\tilde{\delta}(z_{12})\Big(1+\frac{2\bar{\theta}_1\bar{\theta}_2}{\bar{z}_{12}}\Big),~~\tilde{\delta}(z_{12})\equiv 2\pi\delta^{(2)}(z_{12}),
\ee
where $ \p_{z_1} \frac{1}{\bar{z}_{12}}=\tilde{\delta}(z_{12})$     is used.
}
\be\ba\label{crossing1}
&  \p_{z_1} \frac{1}{\bar{Z}_{12}^{2\bar{\Delta}} }=\frac{2\bar{\Delta}}{\bar{Z}_{12}^{2\bar{\Delta}-1}} \tilde{\delta}(z_{12})\Big(1+\frac{2\bar{\theta}_1\bar{\theta}_2}{\bar{z}_{12}}\Big),
\ea\ee
Therefore
\be\ba \label{JP}
&\vev{J(Z)\Phi_1\Phi_2}\\
=&\Big(-\frac{2\Delta}{Z_{12}}\Big( \frac{\theta_{01}}{z_{01}}- \frac{\theta_{02}}{z_{02}}\Big)-\frac{\Delta\theta_{12}}{z_{12}}\Big( \frac{1}{Z_{01}}+\frac{1}{Z_{02}}\Big) +\Delta\Big( \frac{\theta_{01}}{z_{01}^2}+\frac{\theta_{02}}{z_{02}^2}\Big)-2\frac{\bar{\Delta} \theta_{12}}{ z_{01}}(\bar{z}_{12}+\bar{\theta}_1\bar{\theta}_2)\tilde{\delta}(z_{12})   \Big)\vev{\Phi_1\Phi_2}\\
\equiv& P   \vev{\Phi_1\Phi_2}.
\ea\ee
Similarly, the correlator with antiholomorphic component of stress tensor inserted, i.e.  $\vev{\bar{J}(\bar{Z})\Phi_1\Phi_2}$ can be obtained by making the replacement $Z\to \bar{Z},\theta\to \bar{\theta}$ in $P$ defined above, and we denote it as  $\vev{\bar{J}(\bar{Z})\Phi_1\Phi_2}\equiv\bar{P}\vev{\Phi_1\Phi_2}$.

A simplification can be made by noting that to extract $\vev{T(z)\Phi_1\Phi_2}$ from eq.(\ref{JP}), one needs to integrate eq.(\ref{JP}) over $\theta$, and the $\delta$-function term in eq.(\ref{JP}) contains no $\theta$ thus gives no contribution to $\vev{T(z)\Phi_1\Phi_2}$. In view of this point, we will neglect the $\delta$-function terms in both $P$ and $\bar{P}$ hereafter.

Having obtained $\vev{\bar{J}(\bar{Z})\Phi_1\Phi_2}$ we are in position to consider $\vev{J(Z)\bar{J}(\bar{Z})\Phi_1\Phi_2}$, which follows as
\be\ba \label{JJ2pt}
&\vev{J \bar{J}\Phi_1\Phi_2}= \sum_{i=1}^2\Big(\frac{\theta_{0i}}{Z_{0i}}\p_{z_i}+\frac{1}{2Z_{0i}}D_i+\frac{\Delta_i\theta_{0i}}{Z^2_{0i}} \Big)\vev{\bar{J}(\bar{Z})\Phi_1\Phi_2}\equiv(G+F)\vev{\bar{J}(\bar{Z})\Phi_1\Phi_2},
\ea\ee
where we have denoted the terms involving derivatives as $G$, and the remaining terms as $F$ in the second step for later convenience, namely
\be \label{GF}
G= \sum_{i=1}^n \frac{\theta_{0i}}{Z_{0i}}\p_{z_i}+\frac{1}{2Z_{0i}}D_i  ,~~F =\sum_{i=1}^n \frac{\Delta_i\theta_{0i}}{Z^2_{0i}}
\ee
with $n=2$ in the present case (For detailed calculation of eq.(\ref{JJ2pt}), please refer to the appendix \ref{details1} ).

 Finally we obtain the first order integrals as
\be
\ba\label{intJJ}
&\int d^2z d\theta d\bar{\theta}\vev{J(Z)\bar{J}(\bar{Z})\Phi_1(Z_1,\bar{Z}_1)\Phi_n(Z_2,\bar{Z}_2)}/\vev{\Phi_1(Z_1,\bar{Z}_1)\Phi_n(Z_2,\bar{Z}_2)}\\
=&\Delta\bar{\Delta}\int d^2z d\theta d\bar{\theta}\Big[\Big(-\frac{2 }{Z_{12}}\Big( \frac{\theta_{01}}{z_{01}}- \frac{\theta_{02}}{z_{02}}\Big)-\frac{ \theta_{12}}{Z_{12}}\Big( \frac{1}{Z_{01}}+\frac{1}{Z_{02}}\Big) + \Big( \frac{\theta_{01}}{z_{01}^2}+\frac{\theta_{02}}{z_{02}^2}\Big)\Big) \\&
\times \Big(-\frac{2 }{\bar{Z}_{12}}\Big( \frac{\bar{\theta}_{01}}{\bar{z}_{01}}- \frac{\bar{\theta}_{02}}{\bar{z}_{02}}\Big)-\frac{ \bar{\theta}_{12}}{\bar{Z}_{12}}\Big( \frac{1}{\bar{Z}_{01}}+\frac{1}{\bar{Z}_{02}}\Big) + \Big( \frac{\bar{\theta}_{01}}{\bar{z}_{01}^2}+\frac{\bar{\theta}_{02}}{\bar{z}_{02}^2}\Big) \Big)\Big].
\ea
\ee
Expanding the integrand, there will be nine terms. We will consider the first term here and list the remaining eight terms in the appendix \ref{appb}. These integrals can be explicitly performed by employing dimensional regularization which is discussed in section \ref{DR}. More concretely, the first term is (Consider the case $\Delta=\bar{\Delta}$)
\footnote{
Useful relations
\be
\int d\theta \frac{\theta_{01}}{Z_{01}}= \frac{1}{z_{01}},~~\int d\theta \frac{1}{Z_{01}}=\frac{ \theta_1}{z_{01}^2},~~\int d\th d\bar{\th}\bar{\th}\th =1.
\ee
}
\be
\ba\label{T11}
T_{11}\equiv&\int d^2z d\theta d\bar{\theta}\frac{4\Delta^2}{Z_{12}\bar{Z}_{12}}\Big( \frac{\theta_{01}}{z_{01}}- \frac{\theta_{02}}{z_{02}}\Big)\Big( \frac{\bar{\theta}_{01}}{\bar{z}_{01}}- \frac{\bar{\theta}_{02}}{\bar{z}_{02}}\Big)\\
=&-\frac{4\Delta^2}{Z_{12}\bar{Z}_{12}}\int d^2z  \Big( \frac{1}{z_{01}}- \frac{1}{z_{02}}\Big) \Big( \frac{1}{\bar{z}_{01}}- \frac{1}{\bar{z}_{02}}\Big)\\
 =&-\frac{4\Delta^2}{Z_{12}\bar{Z}_{12}}(\CI_{11}(z_1,\bar{z}_1)+\CI_{11}(z_2,\bar{z}_2)-\CI_{11}(z_1,\bar{z}_2)-\CI_{11}(z_2,\bar{z}_1)) \\
=&-\frac{4\Delta^2}{Z_{12}\bar{Z}_{12}}2\pi\Big(-\frac{2}{\epsilon}+\ln|z_{12}|^2+\gamma+\ln\pi\Big),
\ea
\ee
where in the second step we used the notation introduced in eq.(\ref{eq:integrals}), and $\gamma$ is Euler constant and $\epsilon$ is a infinitesimal constant coming from dimensional regularization\footnote{Also $T_{11}$ can be evaluated in an alternatively way as
\be
\ba
T_{11}=&-\frac{4\Delta^2}{Z_{12}\bar{Z}_{12}}\int d^2z  \Big( \frac{1}{z_{01}}- \frac{1}{z_{02}}\Big) \Big( \frac{1}{\bar{z}_{01}}- \frac{1}{\bar{z}_{02}}\Big)\\
=&-\frac{4\Delta^2|z_{12}|^2}{Z_{12}\bar{Z}_{12}}\int d^2z\frac{1}{|z_{01}|^2|z_{02}|^2} \\
=&-\frac{4\Delta^2|z_{12}|^2}{Z_{12}\bar{Z}_{12}} \CI_{1111}(z_1,z_2,\bar{z}_1,\bar{z}_2)\\
=&-\frac{4\Delta^2}{Z_{12}\bar{Z}_{12}}2\pi\Big(-\frac{2}{\epsilon}+\ln|z_{12}|^2+\gamma+\log\pi+ \CO(\epsilon)\Big),
\ea
\ee
which is equal to result in eq.(\ref{T11}). The integral in the last step was computed in \cite{He:2019vzf}.}.
Here the integral
\be
\CI_{11}(z_i,\bar{z}_j)\equiv\int d^2z\frac{1}{z_{0i}\bar{z}_{0j}}
\ee
 is computed in setion \ref{DR}, and we only quote the results in the last line of eq.(\ref{T11}).  

Finally putting together the results of the nine integrals will lead to
\be
\ba
&\frac{1}{\vev{\Phi_1(Z_1,\bar{Z}_1)\Phi_2(Z_2,\bar{Z}_2)}}\int d^2z d\theta d\bar{\theta}\vev{J(Z)\bar{J}(\bar{Z})\Phi_1(Z_1,\bar{Z}_1)\Phi_2(Z_2,\bar{Z}_2)}\\
=&-\frac{4\pi\Delta^2}{Z_{12}\bar{Z}_{12}} \Big(-\frac{4}{\epsilon}+2\ln|z_{12}|^2+2\gamma+2\ln\pi-2\Big).
\ea
\ee
In principle by setting $\theta_{1,2}\to 0$, one can get the results for bosonic CFT, which is
\be \label{susy2}
-\frac{4\pi\Delta^2}{|z_{12}|^2} \Big(-\frac{4}{\epsilon}+2\ln|z_{12}|^2+2\gamma+2\ln\pi-2\Big).
\ee
Comparing this with the CFT results given in eq.(8) in \cite{He:2019vzf} as
\be \label{cft2}
-\frac{4\pi\Delta^2}{|z_{12}|^2} \Big(-\frac{4}{\epsilon}+2\ln|z_{12}|^2+2\gamma+2\ln\pi-5\Big).
\ee
One can find that only  the last constant is different in eq.(\ref{susy2}) and eq.(\ref{cft2}). This difference can be understood from the way we performing the integrals. On one hand, we can use dimensional regularization to evaluate the integral directly
\be
\int d^2z\frac{|z_{12}|^4}{|z_{01}|^4|z_{02}|^4}=-\frac{4\pi\Delta^2}{|z_{12}|^2} \Big(-\frac{4}{\epsilon}+2\ln|z_{12}|^2+2\gamma+2\ln\pi-5\Big),
\ee
which will result in eq.(\ref{cft2}).
On the other hand, we can compute the above integral in an indirect way as we did at the beginning, i.e., expanding the integrand into several terms as below, then using dimensional regularization to compute each integral, finally adding up the contribution of individual term
\be \ba
 \int d^2z\frac{|z_{12}|^4}{|z_{01}|^4|z_{02}|^4}=&\int d^2z\Big(\frac{1}{z^2_{01}}+\frac{1}{z^2_{02}}-\frac{2}{z_{01}z_{02}}\Big)\Big(\frac{1}{\bar{z}^2_{01}}+\frac{1}{\bar{z}^2_{02}}-\frac{2}{\bar{z}_{01}\bar{z}_{02}}\Big)\\
=&-\frac{4\pi\Delta^2}{|z_{12}|^2} \Big(-\frac{4}{\epsilon}+2\ln|z_{12}|^2+2\gamma+2\ln\pi-2\Big),
\ea\ee
which leads to eq.(\ref{susy2}). Nevertheless, the  difference between eq.(\ref{susy2}) and eq.(\ref{cft2}) can be eliminated by  redefining  $\epsilon$.
\subsection{3-point correlators}
The general form of three-point correlators can be written as
\be
\vev{\Phi(Z_1,\bar{Z}_1)\Phi(Z_2,\bar{Z}_2)\Phi(Z_3,\bar{Z}_3)}=cO_3 \bar{O}_3e^{a\theta_{123}\bar{\theta}_{123}},
\ee
where $a,c$ are two undetermined constants and for later convenience we  denote
\be
O_3=\prod_{i<j=1}^3 \frac{1}{Z_{ij}^{\Delta_{ij}}},~~\bar{O}_3=\prod_{i<j=1}^3 \frac{1}{ \bar{Z}_{ij}^{\bar{\Delta}_{ij}}}.
\ee
As discussed in 2-point correlators in the previous subsection, we first consider the correlator $
\vev{J\Phi_1\Phi_2\Phi_3} $ which can be calculated by using the definition of $G,F$ in eq.(\ref{GF}) as follows
\be\ba \label{J31}
&(G+F)O_3\bar{O}_3e^{a\theta_{123}\bar{\theta}_{123}}\\
=&FO_3\bar{O}_3e^{a\theta_{123}\bar{\theta}_{123}}+[G(O_3\bar{O}_3)]e^{a\theta_{123}\bar{\theta}_{123}}+O_3\bar{O}_3[Ge^{a\theta_{123}\bar{\theta}_{123}}]\\
=&(F+P)O_3\bar{O}_3e^{a\theta_{123}\bar{\theta}_{123}}+O_3(G\bar{O}_3)e^{a\theta_{123}\bar{\theta}_{123}}+O_3\bar{O}_3a[(G\theta_{123})\bar{\theta}_{123}-\theta_{123}(G\bar{\theta}_{123})]e^{a\theta_{123}\bar{\theta}_{123}}\\
\to &(F+P+a(G\theta_{123})\bar{\theta}_{123})O_3\bar{O}_3e^{a\theta_{123}\bar{\theta}_{123}},
\ea\ee
where $P$ (defined by $G O_3\equiv PO_3$) turns out to be
\be\ba \label{FP3}
P=\sum_{i,k,k\neq i}\frac{\Delta_{ik}}{Z_{ki}}\Big(\frac{\th_{0i}}{z_{0i}}-\frac{\th_{ki}}{2Z_{0i}}\Big).
\ea\ee
In the last step of eq.(\ref{J31}) we have omitted the "crossing" terms such as $G\bar{O}_3,G\bar{\th}_{123}$ (By crossing terms we mean the terms with holomorphic derivative $\p_z$ acting on antiholomorphic coordinates, or $\p_{\bar{z}}$ acting on holomorphic coordinates, which will result in a $\delta$-function as $\p_z(1/\bar{z})=\tilde{\delta}(z)$. Note that we have encountered crossing term as in eq.(\ref{crossing1}) in the  2-point  correlator case), since these terms will vanish when integrating over $\th$. To be concrete, taking the term  $G\bar{O}_3$ for example
\be\ba \label{cross}
G\bar{O}_3=& -\Big(\frac{\theta_{12}}{z_{01}}\bar{\Delta}_{12}\bar{\theta}_1\bar{\theta}_2\tilde{\delta}(z_{12})+\frac{\theta_{31}}{z_{01}}\bar{\Delta}_{13}\bar{\theta}_3\bar{\theta}_1\tilde{\delta}(z_{31})+\frac{\theta_{23}}{z_{02}}\bar{\Delta}_{23}\bar{\theta}_2\bar{\theta}_3\tilde{\delta}(z_{23}) \Big)
\bar{O}_3\\
&-\frac{\theta_{23}}{2z_{02}}\frac{\bar{\Delta}_{23}\tilde{\delta}(z_{32})}{\bar{Z}^{\bar{\Delta}_{12}}_{12}\bar{Z}^{\bar{\Delta}_{23}-1}_{23}\bar{Z}^{\bar{\Delta}_{13}}_{31}} -\frac{\theta_{31}}{2z_{01}}\frac{\bar{\Delta}_{13}\tilde{\delta}(z_{31})}{\bar{Z}^{\bar{\Delta}_{12} }_{12}\bar{Z}^{\bar{\Delta}_{23}}_{23}\bar{Z}^{\bar{\Delta}_{13}-1}_{31}}-\frac{\theta_{12}}{2z_{01}}\frac{\bar{\Delta}_{12}\tilde{\delta}(z_{12})}{\bar{Z}^{\bar{\Delta}_{12}-1}_{12}\bar{Z}^{\bar{\Delta}_{23}}_{23}\bar{Z}^{\bar{\Delta}_{13} }_{31}},
\ea\ee
thus $\int d\th G\bar{O}_3=0$.

With $\vev{J\Phi_1\Phi_2\Phi_3}$ in hand, we can go on to consider $\vev{J\bar{J}\Phi_1\Phi_2\Phi_3}$
\be\ba \label{JJb31}
&(G+F)(\bar{G}+\bar{F})O_3\bar{O}_3e^{a\theta_{123}\bar{\theta}_{123}}\\
=&(G+F)(\bar{P}+\bar{F}-a\theta_{123}(\bar{G}\bar{\theta}_{123}))O_3\bar{O}_3e^{a\theta_{123}\bar{\theta}_{123}}\\
=& \Big[(F+P+a(G\theta_{123})\bar{\theta}_{123})(\bar{F}+\bar{P}-a\theta_{123}(\bar{G}\bar{\theta}_{123}))\\
&-a(G\theta_{123})(\bar{G}\bar{\theta}_{123})+a\theta_{123}(G(\bar{G}\bar{\theta}_{123}))+G(\bar{F}+\bar{P})\Big]O_3\bar{O}_3  e^{a\theta_{123}\bar{\theta}_{123}}.
\ea\ee
 It is shown in appendix \ref{details2} the last two terms in the last line of above equation will not contribute. Hence from the results in appendix \ref{details2}, the 3-point correlator can be written as 
\be
\ba \label{3ptn1}
&\frac{1}{\vev{\Phi_1\Phi_2\Phi_3}} \int d^2zd\theta d\bar{\theta}\vev{J(Z)\bar{J}(\bar{Z})\Phi_1(Z_1,\bar{Z}_1)\Phi_2(Z_2,\bar{Z}_2)\Phi_3(Z_3,\bar{Z}_3)}\\
=&\int d^2z\Big[\sum_{i,k,k\neq i}\frac{\Delta_{ik}}{Z_{ki}}\Big(\frac{1}{z_{0i}}+\frac{\th_k\th_{i}}{2z^2_{0i}}  \Big)+\sum_i\frac{\Delta_i}{z_{0i}^2}+a\sum_i\Big(\frac{1}{z_{0i}}\p_{z_i}\theta_{123}+\frac{\th_i}{2z^2_{0i}}\p_{\th_i}\th_{123}\Big)\bar{\th}_{123}\Big]\\
&\times (-1)\Big[\sum_{i,k,k\neq i}\frac{\bar{\Delta}_{ik}}{\bar{Z}_{ki}}\Big(\frac{1}{\bar{z}_{0i}}+\frac{\bar{\th}_{k}\bar{\th}_i}{2\bar{z}^2_{0i}}\Big)+\sum_i\frac{\bar{\Delta}_i}{\bar{z}_{0i}^2}+a\th_{123}\sum_i\Big(\frac{1}{\bar{z}_{0i}}\p_{\bar{z}_i}\bar{\theta}_{123}+\frac{\bar{\th}_i}{2\bar{z}^2_{0i}}\p_{\bar{\th}_i}\bar{\th}_{123}\Big)\Big]\\&
-a\sum_i\Big(\frac{1}{z_{0i}}\p_{z_i}\theta_{123}+\frac{\th_i}{2z^2_{0i}}\p_{\th_i}\th_{123}\Big)\times\sum_i\Big(\frac{1}{\bar{z}_{0i}}\p_{\bar{z}_i}\bar{\theta}_{123}+\frac{\bar{\th}_i}{2\bar{z}^2_{0i}}\p_{\bar{\th}_i}\bar{\th}_{123}\Big).
\ea\ee
Let us first consider the terms containing no $a$
\be\ba
-&\int d^2z\sum_{ij}\Big[\sum_{k,k\neq i}\frac{\Delta_{ik}}{Z_{ki}} \frac{1}{z_{0i}}+\frac{1}{z_{0i}^2}\Big(\sum_{k,k\neq i}\frac{\th_k\th_{i}\Delta_{ik}}{2Z_{ki}}+\Delta_i\Big) \Big]
\Big[\sum_{l,l\neq j}\frac{\bar{\Delta}_{lj}}{ \bar{Z}_{lj}} \frac{1}{\bar{z}_{0j}}+\frac{1}{\bar{z}_{0j}^2}\Big(\sum_{l,l\neq j}\frac{\bar{\th}_l\bar{\th}_{j}\bar{\Delta}_{lj}}{2\bar{Z}_{lj}}+\bar{\Delta}_j\Big) \Big] \\
=&-\sum_{ij}\Big[ \CI_{11}(z_i,\bar{z}_j)\sum_{k,k\neq i}\sum_{l,l\neq j}\frac{\Delta_{ik}}{Z_{ki}}\frac{\bar{\Delta}_{lj}}{\bar{Z}_{lj}}+\CI_{22}(z_i,\bar{z}_j)\Big(\sum_{k,k\neq i}\frac{\th_k\th_{i}\Delta_{ik}}{2Z_{ki}}+\Delta_i\Big)\Big(\sum_{l,l\neq j}\frac{\bar{\th}_l\bar{\th}_{j}\bar{\Delta}_{lj}}{2\bar{Z}_{lj}}+\bar{\Delta}_j\Big)\\
&+\CI_{12}(z_i,\bar{z}_j)\sum_{k,k\neq i}\frac{ \Delta_{ik}}{Z_{ki}} \Big(\sum_{l,l\neq j}\frac{\bar{\th}_l\bar{\th}_{j}\bar{\Delta}_{lj}}{2\bar{Z}_{lj}}+\bar{\Delta}_j\Big)+\CI_{21}(z_i,\bar{z}_j)\Big(\sum_{k,k\neq i}\frac{\th_k\th_{i}\Delta_{ik}}{2Z_{ki}}+\Delta_i\Big) \sum_{l,l\neq j}\frac{\bar{\Delta}_{lj}}{ \bar{Z}_{lj}} \Big].
\ea\ee
Next evaluating the $a^1$-terms which contains two parts, the first part is
\be\ba
V_{11}\equiv&-a\sum_{ij}\int d^2z \Big[\sum_{k,k\neq i}\frac{\Delta_{ik}}{2Z_{ki}} \frac{1}{z_{0i}}+\frac{1}{z_{0i}^2}\Big(\sum_{k,k\neq i}\frac{\th_k\th_{i}\Delta_{ik}}{2Z_{ki}}+\Delta_i\Big) \Big] \th_{123} \Big(\frac{1}{\bar{z}_{0j}}\p_{\bar{z}_j}\bar{\theta}_{123}+\frac{\bar{\th}_j}{2\bar{z}^2_{0j}}\p_{\bar{\th}_j}\bar{\th}_{123}\Big)\\
&-(\text{barred}\leftrightarrow \text{unbarred})\\
=&-a\sum_{ij}\Big[\CI_{11}(z_i,\bar{z}_j)\sum_{k,k\neq i}\frac{\Delta_{ik}}{Z_{ki}}\th_{123}\p_{\bar{z}_j}\bar{\theta}_{123}+\CI_{12}(z_i,\bar{z}_j)\sum_{k,k\neq i}\frac{\Delta_{ik}}{2Z_{ki}}\th_{123}\bar{\th}_j\p_{\bar{\th}_j}\bar{\th}_{123}\\
&+\CI_{21}(z_i,\bar{z}_j)\Big(\sum_{k,k\neq i}\frac{\th_k\th_{i}\Delta_{ik}}{2Z_{ki}}+\Delta_i\Big)\th_{123}  \p_{\bar{z}_j}\bar{\theta}_{123}+\frac{1}{2}\CI_{22}(z_i,\bar{z}_j)\Big(\sum_{k,k\neq i}\frac{\th_k\th_{i}\Delta_{ik}}{2Z_{ki}}+\Delta_i\Big)\th_{123}\bar{\th}_j\p_{\bar{\th}_j}\bar{\th}_{123}\Big]\\
&-(\text{barred}\leftrightarrow \text{unbarred}),
\ea\ee
and the second part is
\be\ba
V_{12}\equiv&-a\sum_{ij}\Big(\frac{1}{z_{0i}}\p_{z_i}\theta_{123}+\frac{\th_i}{2z^2_{0i}}\p_{\th_i}\th_{123}\Big)\Big(\frac{1}{\bar{z}_{0j}}\p_{\bar{z}_j}\bar{\theta}_{123}+\frac{\bar{\th}_j}{2\bar{z}^2_{0j}}\p_{\bar{\th}_j}\bar{\th}_{123}\Big)\\
=&-a\sum_{ij}\Big(\CI_{11}(z_i,\bar{z}_j)\p_{z_i}\theta_{123}\p_{\bar{z}_j}\bar{\theta}_{123}+\frac{1}{2}\CI_{12}(z_i,\bar{z}_j)(\p_{z_i}\theta_{123})\bar{\th}_j\p_{\bar{\th}_j}\bar{\th}_{123}+\frac{1}{2}\CI_{21}(z_i,\bar{z}_j)\th_i\p_{\th_i}\th_{123}\p_{\bar{z}_j}\bar{\theta}_{123}\Big),
\ea\ee
As for the $a^2$-term denoted as $V_2$, by observing eq.(\ref{3ptn1}) we find $V_2= -aV_{12}\th_{123} \bar{\th}_{123}$,
thus $V_{12}+V_2=  V_{12} e^{-a\th_{123} \bar{\th}_{123} }$.
In summary, the result for 3-point correlators with $T\bar{T}$ perturbation to first order is
\be
\ba
&\frac{1}{\vev{\Phi_1\Phi_2\Phi_3}} \int d^2zd\theta d\bar{\theta}\vev{J(Z)\bar{J}(\bar{Z})\Phi_1(Z_1,\bar{Z}_1)\Phi_2(Z_2,\bar{Z}_2)\Phi_3(Z_3,\bar{Z}_3)}\\=&-\sum_{i  j}\Big[-\pi\Big(-\frac{2}{\epsilon}+\ln|z_{ij}|^2+\gamma+\ln\pi\Big )\\&\times\sum_{k,k\neq i}\Big(\sum_{l,l\neq j}\frac{\Delta_{ik}}{Z_{ki}}\frac{\bar{\Delta}_{lj}}{\bar{Z}_{lj}}+a\frac{\Delta_{ik}}{Z_{ki}}\th_{123}\p_{\bar{z}_j}\bar{\theta}_{123}+a\frac{\bar{\Delta}_{ik}}{\bar{Z}_{ki}}\p_{z_j}\theta_{123}\bar{\th}_{123}
+a\p_{z_i}\theta_{123}\p_{\bar{z}_j}\bar{\theta}_{123} e^{-a\th_{123}\bar{\th}_{123}}\Big)\\
&+\frac{\pi}{\bar{z}_{ij}}\Big(\frac{a}{2}(\p_{z_i}\theta_{123})\bar{\th}_j\p_{\bar{\th}_j}\bar{\th}_{123}e^{-a\th_{123}\bar{\th}_{123}}+\sum_{k,k\neq i}\frac{\Delta_{ik}}{Z_{ik}} \sum_{l,l\neq j}\frac{\bar{z}_{lj}\bar{\Delta}_{lj}}{2\bar{Z}_{lj}} \\
&+a \sum_{l,l\neq j}\frac{\bar{z}_{lj}\bar{\Delta}_{jl}}{2\bar{Z}_{lj}}  \p_{z_i} \th_{123}\bar{\th}_{123}+a\sum_{k,k\neq i}\frac{\Delta_{ik}}{2Z_{ki}}\th_{123}\bar{\th}_j\p_{\bar{\th}_j}\bar{\th}_{123}\Big)\\
&-\frac{\pi}{z_{ij}}\Big(\frac{a}{2}(\th_i\p_{\th_i}\th_{123})\p_{\bar{z}_j}\bar{\theta}_{123}e^{-a\th_{123}\bar{\th}_{123}}+a \sum_{k,k\neq i}\frac{z_{ki}\Delta_{ik}}{2Z_{ki}} \th_{123}  \p_{\bar{z}_j}\bar{\theta}_{123}
\\&+a\th_i\p_{\th_i}\th_{123}\bar{\th}_{123}\sum_{l,l\neq j}\frac{\bar{\Delta}_{lj}}{2\bar{Z}_{lj}}+ \sum_{k,k\neq i}\frac{z_{ki}\Delta_{ik}}{2Z_{ki}}  \sum_{l,l\neq j}\frac{\bar{\Delta}_{lj}}{\bar{Z}_{lj}} \Big) \Big],
\ea\ee
where the identity $\sum_{k,k\neq i}\bar{\Delta}_{ik}=2\bar{\Delta}_i$ is used to simplify the final expression.

\subsection{$n$-point correlators}
For $n$ point with $n\geq 4$, the undeformed correlator functions take the form as
\be
\vev{\Phi_1(Z_1,\bar{Z}_1)...\Phi_n(Z_n,\bar{Z}_n)}=O_n\bar{O}_n f(U_i,\bar{U}_i,w_k,\bar{w}_k)
\ee
with
\be
O_n=\prod_{i<j}Z_{ij}^{-\Delta_{ij}},~~\bar{O}_n=\prod_{i<j}\bar{Z}_{ij}^{-\bar{\Delta}_{ij}}.
\ee
Assuming all $\Phi_i$ have the same dimension $(\Delta,\bar{\Delta})$, we have
\be
A \equiv \Delta_{ij}=\frac{2\Delta}{n-1},~~\bar{\Delta}\equiv \bar{\Delta}_{ij}=\frac{2\bar{\Delta}}{n-1}.
\ee
Again the crossing terms $\bar{G}Z_{ijkl},\bar{G}\theta_{ijk},\bar{G} O_n$ do not depend on $\th$ and we will not consider these terms below. The computation of correlators with $J$ and $J\bar{J}$ insertion goes in parallel with previous cases, the detailed calculation is presented in appendix \ref{details3}.
Here we only give the results for the $T\bar{T}$ deformed correlator, which takes the form 
\be
\ba
&\frac{1}{\vev{\Phi_1...\Phi_n}} \int d^2zd\theta d\bar{\theta}\vev{J(Z)\bar{J}(\bar{Z})\Phi_1(Z_1,\bar{Z}_1)...\Phi_n(Z_n,\bar{Z}_n)}\\
=&\int d^2z(-1)\sum_{ij}\Big[\sum_{k,k\neq i}\frac{\Delta_{ik}}{Z_{ki}} \frac{1}{z_{0i}}+\frac{1}{z_{0i}^2}\Big(\sum_{k,k\neq i}\frac{\th_k\th_{i}\Delta_{ik}}{2Z_{ki}}+\Delta_i\Big)\Big]\\
&\times \Big[\sum_{l,l\neq j}\frac{\bar{\Delta}_{lj}}{ \bar{Z}_{lj}} \frac{1}{\bar{z}_{0j}}+\frac{1}{\bar{z}_{0j}^2}\Big(\sum_{l,l\neq j}\frac{\bar{\th}_l\bar{\th}_{j}\bar{\Delta}_{lj}}{2\bar{Z}_{lj}}+\bar{\Delta}_j\Big) \Big]\\
&- \sum_{ij}\Big[ \Big( \frac{1}{z_{0i}} \p_{z_i}^{R}f+\frac{\th_i}{2z^2_{0i}}D_i^R f\Big)\frac{1}{f}\Big]   \Big[\sum_{l,l\neq j}\frac{\bar{\Delta}_{lj}}{ \bar{Z}_{lj}} \frac{1}{\bar{z}_{0j}}+\frac{1}{\bar{z}_{0j}^2}\Big(\sum_{l,l\neq j}\frac{\bar{\th}_l\bar{\th}_{j}\bar{\Delta}_{lj}}{2\bar{Z}_{lj}}+\bar{\Delta}_j\Big) \Big]\\
&- \sum_{ij}\Big[\sum_{k,k\neq i}\frac{\Delta_{ik}}{Z_{ki}} \frac{1}{z_{0i}}+\frac{1}{z_{0i}^2}\Big(\sum_{k,k\neq i}\frac{\th_k\th_{i}\Delta_{ik}}{2Z_{ki}}+\Delta_i\Big)\Big]\Big[  \Big( \frac{1}{\bar{z}_{0j}} \p_{\bar{z}_j}^{L}f+\frac{\bar{\th}_j}{2\bar{z}^2_{0j}}\bar{D}_j^L f\Big)\frac{1}{f} \Big]\\
&+\sum_{ij}\Big[ \frac{1}{z_{0i}}  \Big(-\frac{1}{\bar{z}_{0j}} \p_{z_i}^R \p^L_{\bar{z}_j}f-\frac{\bar{\theta}_j}{2\bar{z}^2_{0j}}\p^R_{z_i}\p^L_{\bar{\theta}_j}f\Big)+\frac{\theta_i}{2z_{0i}^2}\Big(-\frac{1}{\bar{z}_{0j}}\p_{\theta_i}^R\p_{\bar{z}_j}^L f +\frac{\bar{\theta}_j}{2\bar{z}^2_{0j}}\p^R_{\theta_i}\p^L_{\bar{\theta}_j}f\Big)\Big].
\ea\ee
Hence using the results for integrals in section \ref{DR}, the final result is
\be\ba \label{N14}
&\frac{1}{\vev{\Phi_1...\Phi_n}} \int d^2zd\theta d\bar{\theta}\vev{J(Z)\bar{J}(\bar{Z})\Phi_1(Z_1,\bar{Z}_1)...\Phi_n(Z_n,\bar{Z}_n)}\\
=&\sum_{i j}\Big[-\pi\Big(-\frac{2}{\epsilon}+\ln|z_{ij}|^2+\gamma+\ln\pi \Big)\\
&\times\Big(-\sum_{k,k\neq i}\frac{\Delta_{ik}}{Z_{ki}} \sum_{l,l\neq j}\frac{\bar{\Delta}_{lj}}{ \bar{Z}_{lj}}-\p^R_{z_i}f \sum_{l,l\neq j}\frac{\bar{\Delta}_{lj}}{ f\bar{Z}_{lj}}-\sum_{k,k\neq i}\frac{\Delta_{ik}}{Z_{ki}}\p^L_{\bar{z}_j}f\frac{1}{f}-\p^R_{z_i}\p^L_{\bar{z}_j}f\frac{1}{f}\Big)\\
&-\frac{\pi}{\bar{z}_{ij}}\Big(\sum_{k,k\neq i}\frac{\Delta_{ik}}{Z_{ki}}
 \sum_{l,l\neq j}\frac{\bar{z}_{lj}\bar{\Delta}_{lj}}{2\bar{Z}_{lj}} +\p^R_{z_i}f \frac{1}{f} \sum_{l,l\neq j}\frac{\bar{z}_{lj}\bar{\Delta}_{lj}}{2\bar{Z}_{lj}} +\sum_{k,k\neq i}\frac{\Delta_{ik}}{2Z_{ki}}\bar{\th}_j\p^L_{\bar{\th}_j}f\frac{1}{f}-\frac{\bar{\th}_j}{2}\p^R_{z_i}\p^L_{\bar{\th}_j}f\frac{1}{f} \Big)\\
&+\frac{\pi}{z_{ij}} \Big(\sum_{k,k\neq i}\frac{z_{ki}\Delta_{ik}}{2Z_{ki}}   \sum_{l,l\neq j}\frac{\bar{\Delta}_{lj}}{ \bar{Z}_{lj}}+\frac{\th_i}{2f}\p_{\th_i}^Rf\sum_{l,l\neq j}\frac{\bar{\Delta}_{lj}}{ \bar{Z}_{lj}}+ \sum_{k,k\neq i}\frac{z_{ki}\Delta_{ik}}{2Z_{ki}}\p^L_{\bar{z}_j}f\frac{1}{f}-\frac{\theta_i}{2 }\p_{\theta_i}^R\p_{\bar{z}_j}^L f \frac{1}{f}\Big)\Big].
\ea\ee
Setting $n=4$, the above results can be used to investigate, for example, the out of time  order correlation function (OTOC). The OTOC is suggested as a diagnostic
of quantum chaos \cite{Shenker:2014cwa,Maldacena:2015waa} since one can extract the Lyapunov exponent corresponding to time evolution form OTOC.  Remarkably, the field theory with Einstein gravity dual is
proposed to exhibit the maximal Lyapunov exponent which measures the growth rate
of the OTOC \cite{ Maldacena:2015waa}. The OTOC of CFT have been considered in \cite{Roberts:2014ifa}, where the OTOC indicates for generic CFT including holographic CFT, the theory have chaotic behavior, but not for integral model such as critical Ising model. Furthermore, the $T\bar{T}$-deformed OTOC for bosonic CFT was investigated in \cite{He:2019vzf}. It was shown that the $T\bar{T}$ deformation does not effect the maximal chaos the for the CFT. In addition, the $T\bar{T}$-deformed integrable model is expected still integrable \cite{He:2019vzf}. This is compatible with the fact that the $T\bar{T}$ deformation does not effect the integrability of the system. Based on these developments, it is will be interesting to investigate the OTOC for $T\bar{T}$-deformed supersymmetric CFT here, we first write the superfield as
\be
\Phi(Z,\bar{Z})=\phi +\th \psi_1+\bar{\th}\psi_2+\th\bar{\th}f
\ee
and its conjugate
\be
\Phi(Z,\bar{Z})^\dagger=\phi^\dagger-\th \psi^\dagger_2-\bar{\th}\psi_1^\dagger+\th\bar{\th}f^\dagger.
\ee
As an example consider the OTOC involving two fields $\phi,\psi_1$, from (45) in \cite{He:2019vzf}, at first order one of the four-point functions needed to compute is 
\be \ba\label{otoc11}
&\vev{\phi(z_1,\bar{z}_1)\phi^\dagger(z_2,\bar{z}_2)\psi_1(z_3,\bar{z}_3)\psi_1^\dagger(z_4,\bar{z}_4)}_{\l}\\
=&-\int d\th_3d\bar{\th}_4 \int d^2zd\theta d\bar{\theta}\vev{J(Z)\bar{J}(\bar{Z})\Phi_1(Z_1,\bar{Z}_1)\Phi^\dagger(Z_2,\bar{Z}_2)\Phi(Z_3,\bar{Z}_3)\Phi^\dagger(Z_4,\bar{Z}_4)}|_{\th_1=\bar{\th}_1=\th_2=\bar{\th}_2=\th_4=\bar{\th}_3=0}\\
=&-\int d\th_3\bar{\th}_4\Big\{\sum_{i\neq j}\Big[-\pi\Big(-\frac{2}{\epsilon}+\ln|z_{ij}|^2+\gamma+\ln\pi \Big)
\\&\times\Big(-\sum_{k,k\neq i}\frac{\Delta_{ik}}{z_{ki}} \sum_{l,l\neq j}\frac{\bar{\Delta}_{lj}}{ \bar{z}_{lj}}f-\p^R_{z_i}f \sum_{l,l\neq j}\frac{\bar{\Delta}_{lj}}{  \bar{z}_{lj}}-\sum_{k,k\neq i}\frac{\Delta_{ik}}{z_{ki}}\p^L_{\bar{z}_j}f-\p^L_{\bar{z}_j}\p^R_{z_i}f \Big)\\
&-\frac{\pi}{\bar{z}_{ij}}\Big(\sum_{k,k\neq i}\frac{\Delta_{ik}}{z_{ki}}
  \bar{\Delta}_j f+\p^R_{z_i}f  \bar{\Delta}_j+\delta_{j4}\sum_{k,k\neq i}\frac{\Delta_{ik}}{2z_{ki}}\bar{\th}_j\p^L_{\bar{\th}_j}f -\delta_{j4}\frac{\bar{\th}_j}{2}\p^R_{z_i}\p^L_{\bar{\th}_j}f \Big)\\
&+\frac{\pi}{z_{ij}}\Big(  \Delta_i \sum_{l,l\neq j}\frac{\bar{\Delta}_{lj}}{ \bar{z}_{lj}}f+\delta_{i3}\frac{\th_i}{2}\p_{\th_i}^Rf\sum_{l,l\neq j}\frac{\bar{\Delta}_{lj}}{ \bar{z}_{lj}}+\Delta_i\p^L_{\bar{z}_j}f -\delta_{i3}\frac{\theta_i}{2 }\p_{\theta_i}^R\p_{\bar{z}_j}^L f  \Big)\Big]\\
&\times\prod_{i<j}z_{ij}^{-\Delta_{ij}}\bar{z}_{ij}^{-\bar{\Delta}_{ij}}\Big\}|_{\th_1=\bar{\th}_1=\th_2=\bar{\th}_2=\th_4=\bar{\th}_3=0},
\ea\ee
where in the integrand, we can replace $Z_{ij}\to z_{ij},\bar{Z}_{ij}\to \bar{z}_{ij}$. In the bosonic CFT, four-point correlators can be expressed as conformal blocks whose universal properties are known in some cases, thus the OTOC can be computed \cite{Roberts:2014ifa}, while in eq.(\ref{otoc11}) the function $f$ is unknown in general. Thus it is more difficult to compute OTOC here.

\section{$\CN$=(2,2) superconformal symmetry }
For (2,2) superconformal symmetry, the coordinates on superspace is divided into holomorphic  $Z=(z,\theta,\bar{\theta})$ and antiholomorphic part $\tilde{Z}=(\bar{z},\tt,\btt)$ respectively. In parallel with the situation in (1,1) case, (2,2) superconformal group is a direct product of (2,0) and (0,2) superconformal group which acts on $Z$ and $\tilde{Z}$ respectively. Thus we will only write out the holomorphic coordinates explicitly hereafter. For holomorphic part the covariant derivatives are \cite{West:1990tg,Blumenhagen:2009zz,DiVecchia:1985ief,DiVecchia:1984nyg,Kiritsis:1987np}
\be
D=\p_{\theta}+\bar{\theta}\p_z,~~\bar{D}=\p_{\bar{\theta}}+\theta\p_z,
\ee
which satisfy $D^2=\bar{D}^2=0,\{D,\bar{D}\}=2\p_z$.
The energy momentum superfield is
\be \ba
J(Z)=j(z)+i\theta\bar{G}(z)+i\bar{\theta}G(z)+2\theta\bar{\theta}T(z),
\ea\ee
and similar for $\bar{J}(\bar{Z})$. Here  $T(z)$ is stress tensor of the theory, and $G(z),\bar{G}(z)$ are two supersymmetric generators, $j(z)$ corresponds to the U(1) symmetry of rotation of the two SUSY charges.

Super-analytic transformation can be defined via the transformation law of covariant derivatives as
\be
D=(D\theta')D',~~\bar{D}=(\bar{D}\bar{\theta}')\bar{D}'.
\ee
Superconformal primary fields are defined such that under super-analytic transformation they transform  as
\be
\Phi(Z)=(D\theta')^{\Delta+Q/2}(\bar{D}\bar{\theta}')^{\Delta-Q/2}\Phi'(Z'),
\ee
where $\Delta,J$ are the dimension and charge of $\Phi$ respectively. The OPE between energy momentum superfield $J(Z)$ and primary superfield have been considered in \cite{DiVecchia:1985ief,Kiritsis:1987np}
\be
J(Z_1)\Phi(Z_2)=2\Delta\frac{\theta_{12}\bar{\theta}_{12}}{Z^2_{12}}\Phi(Z_2)+2\frac{\theta_{12}\bar{\theta}_{12}}{Z_{12}}\p_{z_2}\Phi(Z_2)+\frac{\theta_{12}}{Z_{12}}D\Phi(Z_2)-\frac{\bar{\theta}_{12}}{Z_{12}}\bar{D}\Phi(Z_2)+Q\frac{\Phi(Z_2)}{Z_{12}},
\ee
where $Z_{12}=z_{12}-\theta_1\bar{\theta}_2-\bar{\theta}_1\theta_2$ (also $\tilde{Z}_{12}=\bar{z}_{12}-\tt_1\btt_2-\btt_1\tt_2$).
In analogy with (1,1) case in the previous section, from this OPE, we can get the Ward identity as
\footnote{For the $N=2$ Super-Cauchy theorem see \cite{DiVecchia:1985ief}}
\be\ba \label{ward22}
&\vev{J(Z_0)\Phi_1(Z_1,\tilde{Z}_1)...\Phi_n(Z_n,\tilde{Z}_n)}\\
=&\sum_{i=1}^n \Big(2\Delta_i\frac{\theta_{0i}\bar{\theta}_{0i}}{Z^2_{0i}}+2\frac{\theta_{0i}\bar{\theta}_{0i}}{Z_{0i}}\p_{z_i}+\frac{\theta_{0i}}{Z_{0i}}D_i-\frac{\bar{\theta}_{0i}}{Z_{0i}}\bar{D}_i+\frac{ Q_i}{Z_{0i}}\Big) \vev{\Phi_1(Z_1,\tilde{Z}_1)...\Phi_n(Z_n,\tilde{Z}_n)}.
\ea\ee
In NS sector the $n$-pt correlators on the right hand side of eq.(\ref{ward22}) are constrained by the Ward identity corresponding to global superconformal Osp(2$|$2) transformation \cite{Kiritsis:1987np}.
When $n=2$, the correlator is fixed as
\be \label{2pt2}
\vev{\Phi(Z_1,\tilde{Z}_1)\Phi_n(Z_n,\tilde{Z}_n)}=\frac{1}{Z_{12}^{2\Delta}\tilde{Z}_{12}^{2\bar{\Delta}}}e^{Q_2\frac{\theta_{12}\bar{\theta}_{12}}{Z_{12}}}e^{\bar{Q}_2\frac{\tt_{12}\bar{\tt}_{12}}{\tilde{Z}_{12}}},
\ee
where $\Delta_1=\Delta_2,Q_1+Q_2=0$ and similar for $\bar{\Delta},\bar{Q}$. Note that we have written out the antiholomorphic part explicitly.

For $n=3$ the correlators take  the  form
\be \ba \label{3pt2}
 \vev{\Phi_1(Z_1,\tilde{Z}_1)\Phi_2(Z_2,\tilde{Z}_2)\Phi_3(Z_3,\tilde{Z}_3)}&=\Big(\prod_{i<j}^3  Z_{ij}^{-\Delta_{ij}}\Big) \exp\Big(\sum_{i<j} A_{ij}\frac{\theta_{ij}\bar{\th}_{ij}}{Z_{ij}}\Big)\delta_{Q_1+Q_2+Q_3,0}\\
&\times\Big(\prod_{i<j}^3  \tilde{Z}_{ij}^{-\bar{\Delta}_{ij}}\Big)  \exp\Big(\sum_{i<j} \bar{A}_{ij}\frac{\tt_{ij}\btt_{ij}}{\tilde{Z}_{ij}}\Big)\delta_{\bar{Q}_1+\bar{Q}_2+\bar{Q}_3,0}
\ea\ee
with $A_{ij}=-A_{ji},\sum_{j=1,j\neq i}^3A_{ij}=-Q_i$, and similar for the $\bar{A}_{ij},\bar{Q}_i$. Note that not all $A_{ij}$ are fixed, this is because for three-point case there are nine coordinates $(z_i,\theta_i,\bar{\theta}_i),i=1,2,3$, and eight generators for osp(2$|$2), thus there remains one degree of freedom which corresponds to the invariant quantity
\be
R_{123}=\frac{\th_{12}\bar{\th}_{12}}{Z_{12}}+\frac{\th_{31}\bar{\th}_{31}}{Z_{31}}+\frac{\th_{23}\bar{\th}_{23}}{Z_{23}}
\ee
with $R_{123}^2=0$.

The $n$-pt correlators can be fixed by Ward identity up to an undetermined function
\begin{equation}
\begin{aligned}\label{npt2}
&\vev{\Phi_1(Z_1,\tilde{Z}_1)...\Phi_2(Z_n,\tilde{Z}_n)}\\
=& \Big(\prod_{i<j}^n\frac{1}{Z_{ij}^{\Delta_{ij}}}  \frac{1}{\tilde{Z}_{ij}^{\bar{\Delta}_{ij}}}\Big)\exp\Big(\sum_{i<j}^n \bar{A}_{ij}\frac{\tilde{\theta}_{ij}\btt_{ij}}{\tilde{Z}_{ij}}\Big)\exp \Big(\sum_{i<j}^n A_{ij}\frac{\theta_{ij}\bar{\theta}_{ij}}{Z_{ij}}\Big)\\
&\times f(x_1,x_2,...,x_{3n-8},\bar{x}_1,\bar{x}_2,...,\bar{x}_{3n-8})\delta_{\sum_i Q_i,0} \delta_{\sum_i \bar{Q}_i,0},\\
& A_{ij}=-A_{ji},~\Delta_{ij}=\Delta_{ji},~\sum_{j,j\neq i}A_{ij}=-Q_i,~\sum_{j,j\neq i}\Delta_{ij}=2\Delta_i,\end{aligned}
\end{equation}
where $x_i$ is Osp(2$|$2) invariant variables which may be  either $R_{ijk}$ or $Z_{ijkl}$
\be\ba
R_{ijk}=\frac{\th_{ij}\bar{\th}_{ij}}{Z_{ij}}+\frac{\th_{jk}\bar{\th}_{jk}}{Z_{jk}}
+\frac{\th_{ki}\bar{\th}_{ki}}{Z_{ki}},~~
Z_{ijkl}=\frac{Z_{ij}Z_{kl}}{Z_{li}Z_{jk}}.
\ea\ee
It should be pointed out that only
$3n-8$ variables out of all  $R_{ijk},Z_{ijkl}$ are independent.

In parallel with (1,1) case, we can now define $T\bar{T}$ deformed correlators for (2,2) case. The variation of action under $T\bar{T}$ deformation can be constructed as
\be
\delta S=\l\int d^2zT\bar{T}(z)= \l\int d^2z\int d\theta d\bar{\th} d\tt\btt J(Z)\bar{J}(\bar{Z}),
\ee
Also up to first order the $n$-pt correlators is
\be \label{TTbar22}
-\l\int d^2z\int d\theta d\bar{\theta}d\tt d\btt\vev{J(Z)\bar{J}(\bar{Z})\Phi(Z_1,\bar{Z}_1)...\Phi(Z_n,\bar{Z}_n)}.
\ee
In the following subsections we will consider eq.(\ref{TTbar22}) with $n=2,3$ and $n\geq 4$.

\subsection{2-point correlators}
Up to a constant prefactor, the 2-point correlators have following form
\be
\vev{\Phi_1(Z_1,\tilde{Z}_1)\Phi_2(Z_2,\tilde{Z}_2)}=\frac{1}{Z_{12}^{2\Delta}\tilde{Z}_{12}^{2\bar{\Delta}}}e^{Q_2\frac{\theta_{12}\bar{\theta}_{12}}{Z_{12}}}e^{\bar{Q}_2\frac{\tt_{12}\bar{\tt}_{12}}{\tilde{Z}_{12}}}.
\ee
To obtain $T\bar{T}$ deformed correlators, first considering correlators only with the holomorphic component of stress tensor inserted, from eq.(\ref{ward22}), this is
\be \label{Jphi2}
\vev{J\Phi_1\Phi_2}\equiv(F+G)\vev{\Phi_1\Phi_2},
\ee
where for later convenience we introduced $G,F$ such that $G$ contains derivatives and $F$ does not
\be
G=\sum_{i=1}^n \Big( 2\frac{\theta_{0i}\bar{\theta}_{0i}}{Z_{0i}}\p_{z_i}+\frac{\theta_{0i}}{Z_{0i}}D_i-\frac{\bar{\theta}_{0i}}{Z_{0i}}\bar{D}_i \Big),~~
F=\sum_{i=1}^n \Big(2\Delta_i\frac{\theta_{0i}\bar{\theta}_{0i}}{Z^2_{0i}} +\frac{ Q_i}{Z_{0i}}\Big).
\ee
Having obtained eq.(\ref{Jphi2}), we can go on to consider $J\bar{J}$ inserted correlator which is
\be \label{Jphi22}
\vev{J\bar{J}\Phi_1\Phi_2}=(F+G)(\tilde{F}+\tilde{G})\vev{\Phi_1\Phi_2}
\ee Both eq.(\ref{Jphi2}) and eq.(\ref{Jphi22}) will computed in appendix \ref{details4}. One can note that the procedure is similar to the case $\mathcal{N}=(1,1)$ cases.
 Finally, we get the first order $T\bar{T}$ deformation of 2-point correlators
\be\ba
&\frac{1}{\vev{\Phi_1\Phi_2}}\int d^2zd\th d\bar{\th}d\tt d\btt \vev{J\bar{J}\Phi_1\Phi_2}\\
=&\int d^2zd\th d\bar{\th}d\tt d\btt (F+P)(\tilde{F}+\tilde{P}) \\
=&\int d^2z \Big[-2\Delta\Big(\frac{1}{z_{01}^2}+\frac{1}{z_{02}^2}\Big)-2Q_2\Big(\frac{\bar{\th}_1\th_1}{z_{01}^3}-\frac{\bar{\th}_2\th_2}{z_{02}^3}\Big)+4\Delta\Big(\frac{1}{z_{01}}-\frac{1}{z_{02}}\Big)\frac{1}{Z_{12}}-\Big(\frac{\th_1 }{z^2_{01}}+\frac{\th_{2}}{z^2_{02}}\Big)\frac{2\Delta\bar{\th}_{21}}{Z_{12}}\\
&-\Big(\frac{\bar{\th}_{1}}{z^2_{01}}+\frac{\bar{\th}_{2} }{z^2_{02}}\Big)\frac{2\Delta\th_{21}}{Z_{12}}+2Q_2\frac{\th_{12}\bar{\th}_{12}}{z^2_{12}}\Big(\frac{1}{z_{01}}-\frac{1}{z_{02}}\Big)-Q_2\Big(\frac{\th_{1}}{z^2_{01}}-\frac{\th_{2}}{z^2_{02}}\Big)\frac{\bar{\th}_{12}}{Z_{12}}+Q_2\Big(\frac{\bar{\th}_{1}}{z^2_{01}}-\frac{\bar{\th}_{2}}{z^2_{02}}\Big)\frac{ \th_{12}}{Z_{12}}\Big]\\
&\times\Big[-2\bar{\Delta}\Big(\frac{1}{\bar{z}_{01}^2}+\frac{1}{\bar{z}_{02}^2}\Big)-2\bar{Q}_2\Big(\frac{\btt_1\tt_1}{\bar{z}_{01}^3}-\frac{\btt_2\tt_2}{\bar{z}_{02}^3}\Big)+4\bar{\Delta}\Big(\frac{1}{\bar{z}_{01}}-\frac{1}{\bar{z}_{02}}\Big)\frac{1}{\tilde{Z}_{12}}-\Big(\frac{\tt_1 }{\bar{z}^2_{01}}+\frac{\tt_{2}}{\bar{z}^2_{02}}\Big)\frac{2\bar{\Delta}\btt_{21}}{\tilde{Z}_{12}}\\
&-\Big(\frac{\btt_{1}}{\bar{z}^2_{01}}+\frac{\btt_{2} }{\bar{z}^2_{02}}\Big)\frac{2\bar{\Delta}\tt_{21}}{\tilde{Z}_{12}}+2\bar{Q}_2\frac{\tt_{12}\btt_{12}}{\bar{z}^2_{12}}\Big(\frac{1}{\bar{z}_{01}}-\frac{1}{\bar{z}_{02}}\Big)-\bar{Q}_2\Big(\frac{\tt_{1}}{\bar{z}^2_{01}}-\frac{\tt_{2}}{\bar{z}^2_{02}}\Big)\frac{\btt_{12}}{\tilde{Z}_{12}}+\bar{Q}_2\Big(\frac{\btt_{1}}{\bar{z}^2_{01}}-\frac{\btt_{2}}{\bar{z}^2_{02}}\Big)\frac{ \tt_{12}}{\tilde{Z}_{12}}\Big].
\ea\ee
Further performing the integral over $z$ using dimensional regularization, one can obtain
\be\ba
&\frac{1}{\vev{\Phi_1\Phi_2}}\int d^2zd\th d\bar{\th}d\tt d\btt \vev{J\bar{J}\Phi_1\Phi_2}\\
=& 2\pi\Big(-\frac{2}{\epsilon}+\ln|z_{ij}|^2+\gamma+\ln\pi \Big)\Big(\frac{4\Delta}{Z_{12}}+2Q_2\frac{\th_{12}\bar{\th}_{12}}{z^2_{12}}\Big)\Big(\frac{4\bar{\Delta}}{\tilde{Z}_{12}}+2\bar{Q}_2\frac{\tt_{12}\btt_{12}}{\bar{z}^2_{12}}\Big)\\
&-\frac{\pi}{\bar{z}_{ij}}\Big(\frac{4\Delta}{Z_{12}}+2Q_2\frac{\th_{12}\bar{\th}_{12}}{z^2_{12}}\Big)\Big(\frac{4\bar{\Delta} \bar{z}_{12}}{\tilde{Z}_{12}}+\bar{Q}_2\frac{(\tt_1+\tt_{2})\btt_{12}}{\bar{z}_{12}}-\bar{Q}_2\frac{(\btt_1+\btt_2)\tt_{12}}{\bar{z}_{12}}\Big)\\
&-\frac{\pi}{z_{ij}}\Big(\frac{4\Delta z_{12}}{Z_{12}}+Q_2\frac{\th_{12}(\bar{\th}_1+\bar{\th}_2)}{z_{12}}-Q_2\frac{\bar{\th}_{12}(\th_1+\th_2)}{z_{12}}\Big)\Big(\frac{4\bar{\Delta}}{\tilde{Z}_{12}}+2\bar{Q}_2\frac{\tt_{12}\btt_{12}}{\bar{z}^2_{12}}\Big)\\
&+\frac{\pi}{\bar{z}_{ij}^2}\Big(\frac{4\Delta}{Z_{12}}+2Q_2\frac{\th_{12}\bar{\th}_{12}}{z^2_{12}}\Big)\Big(2\bar{Q}_2\btt_2\tt_2+2\bar{Q}_2\btt_1\tt_1\Big)\\
&+\frac{\pi}{z_{ij}^2}\Big(2Q_2\bar{\th}_1\th_1+2Q_2\bar{\th}_2\th_2\Big)\Big(\frac{4\bar{\Delta}}{\tilde{Z}_{12}}+2\bar{Q}_2\frac{\tt_{12}\btt_{12}}{\bar{z}^2_{12}}\Big).
\ea\ee
\subsection{3-point correlators}
Using the Ward identity, the 3-point correlators take the general form as
\be \ba
 \vev{\Phi_1(Z_1,\tilde{Z}_1)\Phi_2(Z_2,\tilde{Z}_2)\Phi_3(Z_3,\tilde{Z}_3)}&=\Big(\prod_{i<j}^3  Z_{ij}^{-\Delta_{ij}}\Big) \exp\Big(\sum_{i<j} A_{ij}\frac{\theta_{ij}\bar{\th}_{ij}}{Z_{ij}}\Big)\delta_{Q_1+Q_2+Q_3,0}\\
&\times\Big(\prod_{i<j}^3  \tilde{Z}_{ij}^{-\bar{\Delta}_{ij}}\Big)  \exp\Big(\sum_{i<j} \bar{A}_{ij}\frac{\tt_{ij}\btt_{ij}}{\tilde{Z}_{ij}}\Big)\delta_{\bar{Q}_1+\bar{Q}_2+\bar{Q}_3,0}.
\ea\ee
Following the same line as 2-point correlators, we first consider
\be \label{N2J31}
\vev{J\Phi_1\Phi_2\Phi_3}=\vev{G+F}\vev{\Phi_1\Phi_2\Phi_3}.
\ee Based on this equation, we can go further investigate the $J\bar{J}$ insertion
\be\ba \label{JJ2231}
\vev{J\bar{J}\Phi_1\Phi_2\Phi_3}=&(G+F)(\tilde{G}+\tilde{F})\vev{\Phi_1\Phi_2\Phi_3}.
\ea\ee
The detailed computations of eq.(\ref{N2J31}) and eq.(\ref{JJ2231}) are similar to the 2-point case, and are presented in appendix \ref{details5}.
Consequently, the final result for 3-point correlators are
\be\ba
&\frac{1}{\vev{\Phi_1\Phi_2\Phi_3}}\int d^2zd\th d\bar{\th}d\tt d\btt \vev{J\bar{J}\Phi_1\Phi_2\Phi_3}\\
=&\sum_{ij}\Big[ -\pi\Big(-\frac{2}{\epsilon}+\ln|z_{ij}|^2+\gamma+\ln\pi  \Big)\sum_{k,i\neq k}  \Big(\frac{2\Delta_{ki}}{Z_{ki}}+\frac{2\th_{ki}\bar{\th}_{ki}A_{ki}}{z^2_{ki}}\Big)\sum_{l,j\neq l}\Big(  \frac{2\bar{\Delta}_{lj}}{\tilde{Z}_{lj}}+\frac{2\tt_{lj}\btt_{lj}\bar{A}_{lj}}{\bar{z}^2_{lj}}\Big)\\
&+\frac{\pi}{\bar{z}_{ij}}\sum_{k,i\neq k}  \Big(\frac{2\Delta_{ki}}{Z_{ki}}+\frac{2\th_{ki}\bar{\th}_{ki}A_{ki}}{z^2_{ki}}\Big)\sum_{l,j\neq l}\Big(\frac{\bar{z}_{lj}\bar{\Delta}_{lj}}{\tilde{Z}_{lj}}+\bar{A}_{jl} \frac{\tt_j\btt_{jl}-\btt_j\tt_{jl}}{\bar{z}_{jl}}\Big)\\
&-\frac{\pi}{z_{ij}}\sum_{k,i\neq k}  \Big(\frac{z_{ki}\Delta_{ki}}{Z_{ki}}+A_{ik} \frac{\th_i\bar{\th}_{ik}-\bar{\th}_i\th_{ik}}{z_{ik}}\Big)\sum_{l,j\neq l}\Big(  \frac{2\bar{\Delta}_{lj}}{\tilde{Z}_{lj}}+\frac{2\tt_{lj}\btt_{lj}\bar{A}_{lj}}{\bar{z}^2_{lj}}\Big)\\
&-\frac{\pi}{\bar{z}_{ij}^2} \sum_{k,i\neq k}  \Big(\frac{2\Delta_{ki}}{Z_{ki}}+\frac{2\th_{ki}\bar{\th}_{ki}A_{ki}}{z^2_{ki}}\Big) 2\bar{Q}_j\btt_j\th_j - \frac{\pi}{z_{ij}^2} 2Q_i\bar{\th}_i\th_i\sum_{l,j\neq l}\Big(  \frac{2\bar{\Delta}_{lj}}{\tilde{Z}_{lj}}+\frac{2\tt_{lj}\btt_{lj}\bar{A}_{lj}}{\bar{z}^2_{lj}}\Big)\Big].
\ea\ee
\subsection{$n$-point correlators}
The $n$-point function can be fixed by the Ward identity up to an undetermined function
\begin{equation}\label{N24ept}
\begin{aligned}
&\vev{\Phi_1(Z_1,\tilde{Z}_1)...\Phi_n(Z_n,\tilde{Z}_n)}\\
=& \Big(\prod_{i<j}^n\frac{1}{Z_{ij}^{\Delta_{ij}}}  \frac{1}{\tilde{Z}_{ij}^{\bar{\Delta}_{ij}}}\Big)\exp\Big(\sum_{i<j}^n \bar{A}_{ij}\frac{\tilde{\theta}_{ij}\btt_{ij}}{\tilde{Z}_{ij}}\Big)\exp \Big(\sum_{i<j}^n A_{ij}\frac{\theta_{ij}\bar{\theta}_{ij}}{Z_{ij}}\Big)\\
&\times f(x_1,x_2,...,x_{3n-8},\bar{x}_1,\bar{x}_2,...,\bar{x}_{3n-8})\delta_{\sum_i Q_i,0} \delta_{\sum_i \bar{Q}_i,0},
\end{aligned}
\end{equation}
where $x_i$ can be either of the following invariant variables
\be\ba
R_{ijk}=\frac{\th_{ij}\bar{\th}_{ij}}{Z_{ij}}+\frac{\th_{jk}\bar{\th}_{jk}}{Z_{jk}}
+\frac{\th_{ki}\bar{\th}_{ki}}{Z_{ki}},~~
Z_{ijkl}=\frac{Z_{ij}Z_{kl}}{Z_{li}Z_{jk}}.
\ea\ee
Note that the prefactor in front of the function $f$ in eq.(\ref{N24ept}) takes the same form as 3-point correlator, therefore the only difference for $J$ and $J\bar{J}$ inserted correlators from the 3-point case is the effect of function $f$ in eq.(\ref{N24ept}). The main details are included in appendix \ref{details6}.
 After integration eq.(\ref{preint}) explicitly, the final result is
{\small{
\be\ba \label{aint}
&\frac{1}{\vev{\Phi_1...\Phi_n}}\int d^2zd\th d\bar{\th}d\tt d\btt \vev{J\bar{J}\Phi_1...\Phi_n}\\
=&-\sum_{ij}\pi\Big(-\frac{2}{\epsilon}+\ln|z_{ij}|^2+\gamma+\ln\pi\Big)\Big(\sum_{k,i\neq k}  \Big(\frac{2\Delta_{ki}}{Z_{ki}}+\frac{2\th_{ki}\bar{\th}_{ki}A_{ki}}{z^2_{ki}}\Big)\sum_{l,j\neq l}\Big(  \frac{2\bar{\Delta}_{lj}}{\tilde{Z}_{lj}}+\frac{2\tt_{lj}\btt_{lj}\bar{A}_{lj}}{\bar{z}^2_{lj}}\Big)\\
&+2\p^R_{z_i}f\frac{1}{f}\sum_{k,k\neq j}\Big(\frac{2\bar{\Delta}_{kj}}{\tilde{Z}_{kj}}+\frac{2\tt_{kj}\btt_{kj}\bar{A}_{kj}}{\bar{z}^2_{kj}}\Big)+\sum_{k,k\neq j}\Big(\frac{2\Delta_{ki}}{Z_{ki}}+\frac{2\th_{ki}\bar{\th}_{ki}A_{ki}}{z^2_{ki}}\Big)2\p_{\bar{z}_j}^L f\frac{1}{f}+4\p^L_{\bar{z}_j}\p^R_{z_i}f\frac{1}{f}\Big)\\
&+\sum_{ij}\frac{\pi}{\bar{z}_{ij}}\Big(\sum_{k,i\neq k}  \Big(\frac{2\Delta_{ki}}{Z_{ki}}+\frac{2\th_{ki}\bar{\th}_{ki}A_{ki}}{z^2_{ki}}\Big)\sum_{l,j\neq l}\Big(\frac{\bar{z}_{lj}\bar{\Delta}_{lj}}{\tilde{Z}_{lj}}+\bar{A}_{jl} \frac{\tt_j\btt_{jl}-\btt_j\tt_{jl}}{\bar{z}_{jl}}\Big)\\
&+2\p^R_{z_i}f \sum_{k,k\neq j}\Big(\frac{\bar{z}_{kj}\bar{\Delta}_{kj}}{\tilde{Z}_{kj}}+\bar{A}_{jk} \frac{\tt_j\btt_{jk}-\btt_j\tt_{jk}}{\bar{z}_{jk}}
\Big)\frac{1}{f}+\sum_{k,k\neq i}\Big(\frac{2\Delta_{ki}}{Z_{ki}}+\frac{2\th_{ki}\bar{\th}_{ki}A_{ki}}{z^2_{ki}}\Big)(\tt_{j} \p_{\tt_j}^L f +\btt_j\p^L_{\btt_j}f) \frac{1}{f}\\
&+(2\btt_j\p^L_{\btt_j}\p^R_{z_i}f+2\tt_j\p^L_{\tt_j}\p^R_{z_i}f)\frac{1}{f}\Big) -\sum_{ij}\frac{\pi}{z_{ij}}\Big(\sum_{k,i\neq k}  \Big(\frac{z_{ki}\Delta_{ki}}{Z_{ki}}+A_{ik} \frac{\th_i\bar{\th}_{ik}-\bar{\th}_i\th_{ik}}{z_{ik}}\Big)\sum_{l,j\neq l}\Big(  \frac{2\bar{\Delta}_{lj}}{\tilde{Z}_{lj}}+\frac{2\tt_{lj}\btt_{lj}\bar{A}_{lj}}{\bar{z}^2_{lj}}\Big)\\
&+(\th_i\p^R_{\th_i}f+\bar{\th}_i\p^R_{\bar{\th}_j}f) \sum_{k,k\neq j}\Big(\frac{2\bar{\Delta}_{kj}}{\tilde{Z}_{kj}}+\frac{2\tt_{kj}\btt_{kj}\bar{A}_{kj}}{\bar{z}^2_{kj}}\Big)\frac{1}{f}+\sum_{k,k\neq i}\Big(\frac{z_{ki}\Delta_{ki}}{Z_{ki}}+A_{ik} \frac{\th_i\bar{\th}_{ik}-\bar{\th}_i\th_{ik}}{z_{ik}}\Big)2\p_{\bar{z}_j}^L f\frac{1}{f}\\
&+(2 \th_{i} \p_{\th_i}^R\p_{\bar{z}_j}^L f+2 \bar{\th}_{i} \p_{\bar{\th}_i}^R \p_{\bar{z}_j}^L f)\frac{1}{f}\Big) -\sum_{ij}\frac{\pi}{\bar{z}_{ij}^2}\Big(  4\p^R_{z_i}f\frac{1}{f}\bar{Q}_j\btt_j\tt_j+\sum_{k,i\neq k}  \Big(\frac{2\Delta_{ki}}{Z_{ki}}+\frac{2\th_{ki}\bar{\th}_{ki}A_{ki}}{z^2_{ki}}\Big) 2\bar{Q}_j\btt_j\th_j\Big)
\\&
-\sum_{ij}\frac{\pi}{z_{ij}^2}\Big(4Q_i\bar{\th}_i\th_i \p^L_{\bar{z}_j}f\frac{1}{f}+ 2Q_i\bar{\th}_i\th_i\sum_{l,j\neq l}\Big(  \frac{2\bar{\Delta}_{lj}}{\tilde{Z}_{lj}}+\frac{2\tt_{lj}\btt_{lj}\bar{A}_{lj}}{\bar{z}^2_{lj}}\Big)\Big)\\
\ea\ee}}
As an application of last equation, we briefly discuss the 4-point functions that might be useful in the study of the deformed OTOC. The superfield in (2,2) superspace takes the form
\be
\Phi(Z,\tilde{Z})=\phi+\theta \psi_1+...,
\ee
where there are total 16 terms at the right hand side, and we only explicitly write out the first two components since we are only interested in correlators involving $\phi,\psi_1$ as we did in (1,1) case. The conjugated superfield then is
\be
\Phi(Z,\tilde{Z})^\dagger=\phi^\dagger-\bar{\th}\psi_1^\dagger+...
\ee
Thus the following operator appeared in first order perturbation of OTOC\be \ba
&\vev{\phi(z_1,\bar{z}_1)\phi^\dagger(z_2,\bar{z}_2)\psi_1(z_3,\bar{z}_3)\psi_1^\dagger(z_4,\bar{z}_4)}_{\l}\\
=&-\int d\th_3d\bar{\th}_4 \int d^2zd\theta d\bar{\theta}\vev{J(Z)\bar{J}(\tilde{Z})\Phi(Z_1,\tilde{Z}_1)\\&
~~~~~~~~~~~~~~~~\times\Phi^\dagger(Z_2,\tilde{Z}_2)\Phi(Z_3,\tilde{Z}_3)\Phi^\dagger(Z_4,\tilde{Z}_4)}|_{\th_1=\bar{\th}_1=\th_2=\bar{\th}_2=\th_4=\bar{\th}_3=0,\tt_i=\btt_i=0}
\ea\ee
can be computed by utilizing eq.(\ref{aint}).
\section{ Dimensional regularization}\label{DR}
Using Feynman parametrization and dimensional regularization, one can obtain the following basic integral \cite{Guica:2019vnb}
 (Let $z_1\neq z_2$)
\footnote{The notation of integrals is taken the same form as \cite{He:2019vzf}
\be
\label{eq:integrals}
{\cal I}_{a_1,\cdots, a_m, b_1,\cdots, b_n}(z_{i_1},\cdots,z_{i_m}, \bar{z}_{j_1},\cdots, \bar{z}_{j_n})\equiv\int \frac{ d^2z}{(z-z_{i_1})^{a_1}\cdots (z-z_{i_m})^{a_m} (\bar{z}-\bar{z}_{j_1})^{b_1}\cdots (\bar{z}-\bar{z}_{j_n})^{b_n}}.
\ee}
\be \label{CI11}
 \CI_{11}(z_1,\bar{z}_2) =\int d^2z\frac{1}{z_{01}\bar{z}_{02}}=-\pi\Big(-\frac{2}{\epsilon}+\ln|z_{12}|^2+\gamma+\ln\pi\Big)+O(\epsilon)
\ee
with $\epsilon$ being a infinitesimal constant.
Next consider $\CI_{12}(z_1,\bar{z}_2)$ with $ z_1\neq z_2 $
\be\ba\label{CI12}
\int d^2z\frac{1}{z_{01}\bar{z}_{02}^2}&=\int d^2z \frac{\bar{z}_{01}z_{02}^2}{|z_{01}|^2|z_{02}|^4}\\
&=2\int_0^1 du (1-u) \int d^2z\frac{\bar{z}_{01}z_{02}^2 }{(u|z_{01}|^2+(1-u)|z_{02}|^2)^3}\\
&=2\int_0^1 du (1-u) \int d^2y\frac{2uz_{12}|y|^2-(1-u)u^2z^2_{12}\bar{z}_{12} }{(|y|^2+(1-u)u|z_{12}|^2)^3}\\
&=2z_{12}\int_0^1 du u(1-u) \int d^2y\frac{2|y|^2-A^2 }{(|y|^2+A^2)^3}\\
&=2z_{12}\int_0^1 du u(1-u) V_d\int d\rho \rho^{d-1}\frac{2\rho^2-A^2 }{(\rho^2+A^2)^3}\\
&=2z_{12}\int_0^1 du u(1-u) V_dA^{-2}\frac{1}{4}=\frac{\pi}{\bar{z}_{12}},
\ea\ee
where in the last step $d=2$ is set directly since there is no divergence in the integral, and  analytical continuation of the dimension is not required. Here $
V_d=2\pi^{d/2}/\Gamma(d/2)$ is the area of $(d-1)$-sphere with unit radius, also
we denote $A^2=(1-u)u|z_{12}|^2$ and use the coordinates transformation
\be
z=y+uz_1+(1-u)z_2,~~z_{01}=y-(1-u)z_{12},~~z_{02}=y+uz_{12}
\ee
Let us mention that the result in eq.(\ref{CI12}) is consistent with eq.(\ref{CI11}), i.e. they satisfy $\p_{\bar{z}_2} \CI_{11}(z_i,\bar{z}_j)=\CI_{12}(z_i,\bar{z}_j)$.

For $\CI_{22}(z_1,\bar{z}_2)$ with $z_1\neq z_2$, similarly we can obtain
\be\ba
\int d^2z\frac{1}{z_{01}^2\bar{z}_{02}^2}&=\int d^2z \frac{\bar{z}_{01}^2z_{02}^2}{|z_{01}|^4|z_{02}|^4}\\
&=6\int_0^1 du u(1-u) \int d^2y\frac{(\bar{y}-(1-u)\bar{z}_{12})^2(y+uz_{12})^2 }{(|y|^2+(1-u)u|z_{12}|^2)^4}\\
&=6\int_0^1 duu (1-u) \int d^2y\frac{ |y|^4-4|y|^2u(1-u)|z_{12}|^2+(1-u)^2u^2|z_{12}|^4 }{(|y|^2+(1-u)u|z_{12}|^2)^4}\\
&=6\int_0^1 duu (1-u) \int d^2y\frac{ |y|^4-4|y|^2A^2+A^4 }{(|y|^2+A^2)^4}=0.
\ea\ee

In summary, by using dimensional regularization we can obtain the following basic integrals which appear in $\CN=$(1,1) case
\be \ba\label{DR1}
&\CI_{11}(z_i,\bar{z}_j)=-\pi(-\frac{2}{\epsilon}+\ln|z_{ij}|^2+\gamma+\ln\pi+\mathcal{O}(\epsilon)),\\&
\CI_{12}(z_i,\bar{z}_j)=\frac{ \pi}{\bar{z}_{ij}} ,~~\CI_{21}(z_i,\bar{z}_j)=-\frac{ \pi}{z_{ij}} ,~~\CI_{22}(z_i,\bar{z}_j)=0,\\
&\CI_{11}(z_i,\bar{z}_i)=0,~~\CI_{12}(z_i,\bar{z}_i)=0,~~\CI_{22}(z_i,\bar{z}_i)=0,
\ea\ee
where in the last line  the integrals with two points coincide are listed. For these integrals by translation symmetry, we can set $z_i=0$, thus there is no scale in the integrals and we can set these integrals equal zero in dimensional regularization.
Note that the integral $\CI_{22}(z_i,\bar{z}_j)$ is proportional to  a delta function $\delta^{(2)}(z_{ij})$ in (B.7) of \cite{Guica:2019vnb}. However, we will omit this delta function here due to the fact that once we let $z_i=z_j$ in $\CI_{22}(z_i,\bar{z}_j)$,  as mentioned above, by translation symmetry there is no scale in the integral. Thus the term $\delta^{(2)}(z_{ij})$ in (B.7) of \cite{Guica:2019vnb} is simply replaced by zero in eq.(\ref{DR1}).

By using Feynman parametrization, following the same line as above, we can also obtain the integrals needed in the $\CN=(2,2)$ case, which are
\be\ba
&\CI_{13}(z_i,\bar{z}_j)=\frac{\pi}{(\bar{z}_{ij})^2},~~
\CI_{31}(z_i,\bar{z}_j)=\frac{\pi}{(z_{ij})^2},\\
&  \CI_{23}(z_i,\bar{z}_j)=\CI_{32}(z_i,\bar{z}_j)=\CI_{33}(z_i,\bar{z}_j)=0,
\ea\ee
where we also let the integrals with two points coinciding with each other vanish.

\section{Conclusions}
In the present paper we investigated the correlation functions with $T\bar{T}$ deformation for $\CN=(1,1)$ and $\CN=(2,2)$ superconformal field theory perturbatively to the first order of the deformation. This extends previous work on the correlation function from bosonic CFTs \cite{He:2019vzf} to supersymmetric ones. Much like the bosonic CFT, the undeformed 2- and 3-point functions are almost fixed by global superconformal symmetry, while the $n$-point ($n\geq 4$) functions depend on a undetermined function $f$ which depends on the cross ratio. Since we only focus on the first order correction to the correlation function, the superconformal symmetry is still hold approximately. One can make use of superconformal Ward identities to work out the obvious form of  correlation functions with $T\bar{T}$ deformation. We have shown that the correlation function can be expressed by the several basic integrals listed in the last section. As a consequence, these integrals have been done with dimensional regularization in a systematical way. As a possible application, we briefly mentioned the OTOC in the deformed superconformal CFTs. Unlike the bosonic CFTs, due to unknown function $f$ in 4-point functions, one can not directly apply the final correlation function to evaluate OTOC, in superconformal field theory with the deformation. Thus more informations about the function $f$ is needed to study the OTOC in superconformal CFTs with the deformation.

In the present paper we only considered the effect of $T\bar{T}$ deformation on correlation functions perturbatively near the IR conformal fixed point. Since $T\bar{T}$ deformation is believed to be UV complete, it is interesting to study the correlation functions of superconformal theory in the deep UV region as what has been done for the bosonic CFT in \cite{Cardy:2019qao}. Another interesting problem is to study the correlation functions in $\CN=(1,0)$ and $\CN=(2,0)$ theories, which exist for Lorentz signature. Possibly, one can also consider correction of the $J\bar{T}$ deformation to the correlation in supersymmetric CFTs recently studied in \cite{Jiang:2019trm}.
\section*{Acknowledgements}
The authors are grateful to Bin Chen, Hongfei Shu, Rong-Xin Miao, Jiahui Bao, Chao Yu for useful discussions.
SH thanks the Yukawa Institute for Theoretical Physics at Kyoto University.
Discussions during the workshop YITP-T-19-03 "Quantum Information and String Theory 2019"
were useful to complete this work. SH also would like to appreciate the
financial support from Jilin University and Max Planck Partner group.
JRS is supported by the National Natural Science Foundation of China under Grant No.~11675272. The Project is also
funded by China Postdoctoral Science Foundation
(No. 2019M653137).
\appendix
\section{Computation details}
\subsection{$\mathcal{N}=(1,1)$: 2-point case}\label{details1}
 To evaluate the right hand side of eq.(\ref{JJ2pt}), first consider the anticommutator between $P$ and $J=F+G$, noting $P,G,F$ are all  Grassmannian odd
\be \ba
\{J,P\}R&=J(PR)+P(JR)\\
&=FPR +G(PR) +PFR+P(GR)\\
&=FPR+(GP)R-P(GR) +PFR+P(GR)\\
&=(GP)R
\ea\ee
with $R\equiv\vev{\Phi_1\Phi_2}$. Hence we obtain
\be\label{GPb11}
\vev{J\bar{J}\Phi_1\Phi_2}=(P\bar{P}+(G\bar{P}))\vev{\Phi_1\Phi_2},
\ee
where the first term on the right hand side
\be
\ba
P\bar{P}=&\Delta\bar{\Delta}\Big(-\frac{2 }{Z_{12}}\Big( \frac{\theta_{01}}{z_{01}}- \frac{\theta_{02}}{z_{02}}\Big)-\frac{ \theta_{12}}{z_{12}}\Big( \frac{1}{Z_{01}}+\frac{1}{Z_{02}}\Big) + \Big( \frac{\theta_{01}}{z_{01}^2}+\frac{\theta_{02}}{z_{02}^2}\Big)\Big) \\
&\times\Big(-\frac{2}{\bar{Z}_{12}}\Big( \frac{\bar{\theta}_{01}}{\bar{z}_{01}}- \frac{\bar{\theta}_{02}}{\bar{z}_{02}}\Big)-\frac{ \bar{\theta}_{12}}{\bar{z}_{12}}\Big( \frac{1}{\bar{Z}_{01}}+\frac{1}{\bar{Z}_{02}}\Big) + \Big( \frac{\bar{\theta}_{01}}{\bar{z}_{01}^2}+\frac{\bar{\theta}_{02}}{\bar{z}_{02}^2}\Big)  \Big).
\ea\ee
Here we have omitted $\delta$-function terms in both $P$ and $\bar{P}$ as mentioned above. The second term in eq.(\ref{GPb11}) is
\be \ba
G\bar{P}=&\bar{\Delta}\sum_{i}\Big(\frac{\theta_{0i}}{Z_{0i}}\p_{z_i}+\frac{1}{2Z_{0i}}\p_{\theta_i}+\frac{1}{2Z_{0i}}\theta_i\p_{z_i}\Big)\\& \times\Big(-\frac{2}{\bar{Z}_{12}}
\Big( \frac{\bar{\theta}_{01}}{\bar{z}_{01}}- \frac{\bar{\theta}_{02}}{\bar{z}_{02}}\Big)-\frac{ \bar{\theta}_{12}}{\bar{z}_{12}}\Big( \frac{1}{\bar{Z}_{01}}+\frac{1}{\bar{Z}_{02}}\Big) + \Big( \frac{\bar{\theta}_{01}}{\bar{z}_{01}^2}+\frac{\bar{\theta}_{02}}{\bar{z}_{02}^2}\Big)  \Big),
\ea\ee
where the third term in the first bracket, i.e. $\frac{1}{2Z_{0i}}\theta_i\p_{z_i}...=\frac{1}{2z_{0i}}\theta_i\p_{z_i}...$, will vanish after integral over $\int d\theta$, and the second term $\p_{\th_i}\bar{P}=0$ since $\bar{P}$ does not dependent on $\theta_i$. Thus the only term needed to compute is
\be
\ba
 \bar{\Delta}\sum_{i}\Big(\frac{\theta_{0i}}{Z_{0i}}\p_{z_i}\Big)\Big(-\frac{2}{\bar{Z}_{12}}
\Big( \frac{\bar{\theta}_{01}}{\bar{z}_{01}}- \frac{\bar{\theta}_{02}}{\bar{z}_{02}}\Big)-\frac{ \bar{\theta}_{12}}{\bar{z}_{12}}\Big( \frac{1}{\bar{Z}_{01}}+\frac{1}{\bar{Z}_{02}}\Big) + \Big( \frac{\bar{\theta}_{01}}{\bar{z}_{01}^2}+\frac{\bar{\theta}_{02}}{\bar{z}_{02}^2}\Big) \Big).
\ea
\ee
It turns out the contributions from the second and third terms in the second bracket are nonzero after integration $\int d \theta d\bar{\theta}$, which is
\be
\ba\label{intGP}
&\int d^2z\int d \theta d\bar{\theta} G\bar{P}=2\bar{\Delta}\int d^2z\Big(\frac{ \bar{\th}_1\bar{\th}_2}{\bar{z}_{12}} +1\Big)\Big(\frac{\tilde{\delta}^{(2)}(z_{01})}{|z_{01}|^2}+  \frac{\tilde{\delta}^{(2)}(z_{02})}{|z_{02}|^2}   \Big)
\ea
\ee
where we use $\int d^2z\frac{\tilde{\delta}(z_{12})}{z_{0i}}=0$, which can be obtained in polar coordinates.
This term is divergent and it should be dropped, which can be seen as follows. By observing eq.(\ref{intGP}), one find that it only depends on $\bar{\Delta}$ while not on $\Delta$, in other words, this term is not symmetric under the interchange of $\bar{\Delta}$ and $\Delta$. However $\vev{\bar{J} J \Phi_1...}=-\vev{J \bar{J} \Phi_1...}$ should holds (the minus sign appears due to $J(Z)$ is Grassmann odd), which implies the correlator $ \vev{J \bar{J} \Phi_1...}$ should be symmetric under interchange of $\bar{\Delta}$ and $\Delta$. From this reasoning we will drop these terms.
 \subsection{$\mathcal{N}=(1,1)$: 3-point case}\label{details2}
Let us first focus on the last two terms in () which are crossing terms.  After some computation the last term is
\be \label{GPF2}
\int d\theta d\bar{\theta}G(\bar{P}+\bar{F})=-2\sum_{i} \bar{\Delta}_i \frac{\tilde{\delta}^{(2)}(z_{0i})}{|z_{0i}|^2}+\sum_{i,k,i\neq k}\frac{\tilde{\delta}^{(2)}(z_{0i})}{|z_{0i}|^2}\frac{\bar{\th}_k\bar{\th}_i}{\bar{z}_{ki}}\bar{\Delta}_{ik}.
\ee
For the same reason as discussed below eq.(\ref{intGP}), this term should be dropped out.
As for the term $G(\bar{G}\bar{\theta}_{123})$, after employing the anti-commutator
\be \ba
\{G,\bar{G}\}=& \sum_{i}\Big(\frac{ \theta_{0i}}{z_{0i}}+\frac{\theta_i}{2z_{0i}}\Big)(-\tilde{\delta}(z_{0i}))\Big(\bar{\theta}_{0i}+\frac{\bar{\theta}_i}{2}\Big)\bar{\p_i}+\sum_{i}\Big(\frac{ \bar{\theta}_{0i}}{\bar{z}_{0i}}+\frac{\bar{\theta}_i}{2\bar{z}_{0i}}\Big)(-\tilde{\delta}(z_{0i}))\Big(\theta_{0i}+\frac{\theta_i}{2}\Big)\p_i \\
&+ \sum_{i}\Big(\frac{ \theta_{0i}}{z_{0i}}+\frac{\theta_i}{2z_{0i}}\Big)(-\tilde{\delta}(z_{0i}))\Big(1+\frac{2\bar{\theta}\bar{\theta}_i}{\bar{z}_{0i}}\Big)\frac{1}{2}\p_{\bar{\theta}_i}+\sum_{i}\Big(\frac{ \bar{\theta}_{0i}}{\bar{z}_{0i}}+\frac{\bar{\theta}_i}{2\bar{z}_{0i}}\Big)(-\tilde{\delta}(z_{0i}))\Big(1+\frac{2\theta\theta_i}{z_{0i}}\Big)\frac{1}{2}\p_{\theta_i}
\ea\ee
$G(\bar{G}\bar{\theta}_{123})$ can be written as
\be \ba
&G(\bar{G}\bar{\theta}_{123})\to \{G,\bar{G}\} \bar{\theta}_{123}\\
\to& \sum_{i}\Big(\frac{ \theta_{0i}}{z_{0i}}+\frac{\theta_i}{2z_{0i}}\Big)(-\tilde{\delta}(z_{0i}))\Big(1+\frac{2\bar{\theta}\bar{\theta}_i}{\bar{z}_{0i}}\Big)\frac{1}{2}\p_{\bar{\theta}_i}\bar{\theta}_{123}\\
&+ \sum_{i}\Big(\frac{ \theta_{0i}}{z_{0i}}+\frac{\theta_i}{2z_{0i}}\Big)(-\tilde{\delta}(z_{0i}))\Big(\bar{\theta}_{0i}+\frac{\bar{\theta}_i}{2}\Big)\bar{\p_i} \bar{\theta}_{123}
\ea\ee
where the term $G\bar{\theta}_{123}$ is omitted in the first step, and also for $\p_{z_j} \bar{\theta}_{123}$ in the second step since they do not contain $\theta$. Thus finally we get
\be \ba \label{tt1}
\int d^2z d\theta d\bar{\theta} G(\bar{G}\bar{\theta}_{123}) =\sum_{i}\int d^2z \frac{-\tilde{\delta}(z_{0i})\bar{\theta}_i}{|z_{0i}|^2}\p_{\bar{\theta}_i}\bar{\theta}_{123},
\ea\ee
which is also singular and should be dropped. This can be seen by noting that if we interchange the position in $\vev{J(Z)\bar{J}(\bar{Z})\Phi_1...}$, and to consider $\vev{\bar{J}(\bar{Z})J(Z)\Phi_1...}$ we will obtain a term different with eq.(\ref{tt1}) as
\be \ba
\int d^2z d\theta d\bar{\theta} \bar{G}(G\theta_{123}) =-\sum_{i}\int d^2z \frac{-\tilde{\delta}(z_{0i})\theta_i}{|z_{0i}|^2}\p_{\theta_i}\theta_{123}.
\ea\ee
thus the appearance of eq.(\ref{tt1}) implies the identity $\vev{\bar{J}(\bar{Z})J(Z)\Phi_1...}=-\vev{J(Z)\bar{J}(\bar{Z})\Phi_1...}$ does not hold. Thus we must drop the crossing term eq.(\ref{tt1}). From this consideration we will omit all the crossing terms without explicitly pointing out in the following case with $n\geq 4$ point correlation functions.
\subsection{$\mathcal{N}=(1,1)$: $n$-point case}\label{details3}
Now evaluate
\be\ba\label{TOn}
\vev{J\Phi_1...\Phi_n}&=(F+G)O_n\bar{O}_n f=(F+P)O_n\bar{O}_n f+Q O_n\bar{O}_n,
\ea\ee
where $P$ takes the same form as eq.(\ref{FP3}) with summation from $1$ to $n$, and
\be \ba
Q&\equiv(G U_i)\frac{\p f}{\p U_i}+(G w_k)\frac{\p f}{\p w_k}=\sum_{j=1}^n\Big( \frac{\theta_{0j}}{Z_{0j}} \p_{z_j}^{R}f+\frac{1}{2Z_{0j}}D_j^R f\Big),
\ea\ee
where we introduced the notation $\p_{z_j}^R, D^R_j,\p_\theta^R$ which act on $z_i,\theta_i$ but not on $\bar{z}_i,\bar{\theta}_i$, and similarly let $\p_{\bar{z}_j}^L, \bar{D}^L_j,\p_{\bar{\theta}}^L$ act on  $\bar{z}_i,\bar{\theta}_i$ but not on $z_i,\theta_i$ (thus $\p_{z_j}^R(1/\bar{z}_j)=0$).  When inserting $J\bar{J}$, yields
\be \ba \label{TT411}
&(F+G)[(\bar{F}+\bar{P})O_n\bar{O}_n f+\bar{Q}O_n\bar{O}_n]\\
=& (F+P)(\bar{F}+\bar{P})O_n\bar{O}_n f+Q(\bar{F}+\bar{P})O_n\bar{O}_n+(F+P)\bar{Q}O_n\bar{O}_n+(G\bar{Q})O_n\bar{O}_n
\ea\ee
with
\be \ba \label{GQb}
\bar{Q}&=\sum_{j=1}^n\Big(\Big(\frac{\bar{\theta}_{0j}}{\bar{z}_{0j}}+\frac{\bar{\theta}_j}{2\bar{z}_{0j}}\Big)\p^L_{\bar{z}_j} f+\frac{1}{2\bar{Z}_{0j}}\p^L_{\bar{\theta}_j}f\Big).
\ea\ee
Naively the last term in eq.(\ref{TT411}) looks like a crossing term, but this is not the case as can be see below
\be\ba \label{GQ}
 G\bar{Q}=&-\sum_{ij}\Big(\frac{\bar{\theta}_{0j}}{\bar{z}_{0j}}+\frac{\bar{\theta}_j}{2\bar{z}_{0j}}\Big)\Big((\p_{\bar{z}_j}\bar{U}_i)\Big(G\frac{\p f}{\p \bar{U}_i}\Big)-(\p_{\bar{z}_j}\bar{w}_i)\Big(G\frac{\p f}{\p \bar{w}_i}\Big)\Big)\\&-\sum_{ij}\frac{1}{2\bar{Z}_{0j}}\Big((\p_{\bar{\theta}_j}\bar{U}_i)\Big(G\frac{\p f}{\p \bar{U}_i}\Big)-(\p_{\bar{\theta}_j}\bar{w}_i)\Big(G\frac{\p f}{\p \bar{w}_i}\Big)\Big),
\ea\ee
where for example one has
\be\ba
&G\frac{\p f}{\p \bar{U}_i}=\sum_j\Big((GU_j)\frac{\p^2 f}{\p U_j\p \bar{U}_i} +(Gw_j)\frac{\p^2 f}{\p w_j\p \bar{U}_i} \Big)\equiv   G^R\frac{\p f}{\p \bar{U}_i}
\ea\ee
with $G^R$  acting only on $U_j,w_j$ but not on $\bar{U}_j,\bar{w}_j$.
Eventually one can get
\be\ba
\int d\th d\bar{\th} G\bar{Q}=\sum_{ij}\Big[ \frac{1}{z_{0i}}  \Big(-\frac{1}{\bar{z}_{0j}} \p_{z_i}^R \p^L_{\bar{z}_j}f-\frac{\bar{\theta}_j}{2\bar{z}^2_{0j}}\p^R_{z_i}\p^L_{\bar{\theta}_j}f\Big)+\frac{\theta_i}{2z_{0i}^2}\Big(-\frac{1}{\bar{z}_{0j}}\p_{\theta_n}^R\p_{\bar{z}_j}^L f +\frac{\bar{\theta}_j}{2\bar{z}^2_{0j}}\p^R_{\theta_n}\p^L_{\bar{\theta}_j}f\Big)\Big].
\ea\ee
\subsection{$\mathcal{N}=(2,2)$: 2-point case}\label{details4}
To evaluate eq.(\ref{Jphi2}), firstly, let us consider the crossing terms (holomorphic derivatives $\p_{z}$ acting on antiholomorphic coordinates  or vice versa) in eq.(\ref{Jphi2}). In analogy with the $(1,1)$ case, it can be shown that this kind of terms vanish when integrating over $\th,\bar{\th}$, thus it will not contribute to the final results eq.(\ref{TTbar22}). Explicitly, consider the crossing term $G  \frac{1}{\tilde{Z}_{12}^{2\bar{\Delta}}}$,
\footnote{Some useful expressions
\be
\frac{1}{Z_{0i}}=\frac{1}{z_{0i}}+\frac{\theta_0\bar{\theta}_i+\bar{\th}_0\th_i}{z_{0i}^2}+2\frac{\th_0\bar{\th}_i\bar{\th}_0\th_i}{z_{0i}^3},
\ee
\be
\frac{\th_{0i}}{Z_{0i}}=\frac{\th_{0i}}{z_{0i}}-\frac{\theta_0\th_i\bar{\theta}_{0i}}{z^2_{0i}},~~
\frac{\bar{\th}_{0i}}{Z_{0i}}=\frac{ \bar{\th}_{0i}}{z_{0i}}-\frac{\bar{\th}_0\bar{\th}_i\th_{0i}}{z_{0i}^2},~~
\frac{\theta_{0i}\bar{\th}_{0i}}{Z_{0i}}=\frac{\theta_{0i}\bar{\th}_{0i}}{z_{0i}},
\ee
\be
\int d\th d\bar{\th} \bar{\th}\th=1.
\ee
}
\be\ba \label{G1}
G  \frac{1}{\tilde{Z}_{12}^{2\bar{\Delta}}}=&\int d\th d\bar{\th}\sum_i\Big( 2\frac{\theta_{0i}\bar{\theta}_{0i}}{Z_{0i}}\p_{z_i}+\frac{\theta_{0i}}{Z_{0i}}D_i-\frac{\bar{\theta}_{0i}}{Z_{0i}}\bar{D}_i\Big) \frac{1}{\tilde{Z}_{12}^{2\bar{\Delta}}} \\
=&\int d\th d\bar{\th} \Big(\frac{\theta_{01}}{z_{01}}\bar{\th}_1-\frac{\bar{\theta}_{01}}{z_{01}}\th_1
-\frac{\theta_{02}}{z_{02}}\bar{\th}_2+\frac{\bar{\theta}_{02}}{z_{02}}\th_2+2\Big(\frac{\theta_{01}\bar{\th}_{01}}{z_{01}}-\frac{\theta_{02}\bar{\th}_{02}}{z_{02}}\Big)\Big)\frac{2\bar{\Delta}}{\tilde{Z}_{12}^{2\bar{\Delta}-1}}\p_{z_1}\frac{1}{\tilde{Z}_{12}}=0,
\ea\ee
where in the last step we have used
\be \ba \label{delde}
\p_{z_1}\frac{1}{\tilde{Z}_{12}}=-\p_{z_2}\frac{1}{\tilde{Z}_{12}}=\tilde{\delta}(z_{12})\Big(1+2\frac{\tt_1\btt_2+\btt_1\tt_2}{\bar{z}_{12}}+6\frac{\tt_1\btt_2\btt_1\tt_2}{\bar{z}_{12}^2}\Big).
\ea\ee
In the same manner one has $\int d\th d\bar{\th}Ge^{\bar{Q}_2\frac{\tt_{12}\bar{\tt}_{12}}{\tilde{Z}_{12}}}=0$.
Therefore we can derive eq.(\ref{Jphi2}) in the following without considering crossing terms, which is
\be
\vev{J\Phi_1\Phi_2}=(F+G)\vev{\Phi_1\Phi_2}\equiv(F+P)\vev{\Phi_1\Phi_2}
\ee
with \be \ba
F&= 2\Delta\Big(\frac{\theta_{01}\bar{\th}_{01}}{z_{01}^2}+\frac{\theta_{02}\bar{\th}_{02}}{z_{02}^2}\Big)-Q_2\Big(\frac{1}{Z_{01}}-\frac{1}{Z_{02}}\Big)
\ea\ee
and $P$ is defined as (similar for $\tilde{P},\tilde{F}$)
\be\ba
&P\equiv G\vev{\Phi_1\Phi_2}/\vev{\Phi_1\Phi_2}\\
=&-4\Delta\Big(\frac{\theta_{01}\bar{\th}_{01}}{z_{01}}-\frac{\theta_{02}\bar{\th}_{02}}{z_{02}}\Big)\frac{1}{Z_{12}}+\Big(\frac{\th_{01}\bar{\th}_{21}}{Z_{01}}+\frac{\th_{02}\bar{\th}_{21}}{Z_{02}}\Big)\frac{2\Delta}{Z_{12}}\\
&-\Big(\frac{\bar{\th}_{01}\th_{21}}{Z_{01}}+\frac{\bar{\th}_{02}\th_{21}}{Z_{02}}\Big)\frac{2\Delta}{Z_{12}}-2Q_2\frac{\th_{12}\bar{\th}_{12}}{z^2_{12}}\Big(\frac{\theta_{01}\bar{\th}_{01}}{z_{01}}-\frac{\theta_{02}\bar{\th}_{02}}{z_{02}}\Big)\\
&+Q_2\Big(\frac{\th_{01}}{Z_{01}}-\frac{\th_{02}}{Z_{02}}\Big)\frac{\bar{\th}_{12}}{Z_{12}}+Q_2\Big(\frac{\bar{\th}_{01}}{Z_{01}}-\frac{\bar{\th}_{02}}{Z_{02}}\Big)\frac{ \th_{12}}{Z_{12}}.
\ea\ee
Having obtained eq.(\ref{Jphi2}), next we can investigate $\vev{J\bar{J}\Phi_1\Phi_2}$
\be \ba
&(F+G)(\tilde{F}+\tilde{G})\vev{\Phi_1\Phi_2}=(F+G)(\tilde{F}+\tilde{P})\vev{\Phi_1\Phi_2}\\
=&(F+P)(\tilde{F}+\tilde{P})O_2\tilde{O}_2+[G(\tilde{F}+\tilde{P})]\vev{\Phi_1\Phi_2}.
\ea\ee
Note that the last term is also a crossing term which can be dropped by the same reason as discussed around eq.(\ref{GPF2}).
\subsection{$\mathcal{N}=(2,2)$: 3-point case}\label{details5}
It can be shown that the crossing terms in eq.(\ref{N2J31}) do not contribute, i.e.
\be \label{cross2}
\int d\th d\bar{\th}G\Big(\prod_{i<j}^3  \tilde{Z}_{ij}^{-\bar{\Delta}_{ij}}\Big)=0,~~\int d\th d\bar{\th}G \exp\Big(\sum_{i<j} \bar{A}_{ij}\frac{\tt_{ij}\btt_{ij}}{\tilde{Z}_{ij}}\Big)=0.
\ee
Therefore we only need to consider
 \be\ba \label{p1}
G\Big(\prod_{i<j}^3  Z_{ij}^{-\Delta_{ij}}\Big)=&\sum_{i,k,i\neq k}\Big(2\frac{\th_{0k}\bar{\th}_{0k}}{z_{0k}}\frac{\Delta_{ik}}{Z_{ik}}+\frac{\th_{0k}}{Z_{0k}}\frac{\bar{\th}_{ki}\Delta_{ik}}{Z_{ik}}- \frac{\bar{\th}_{0k}}{Z_{0k}}\frac{\th_{ki}\Delta_{ik}}{Z_{ik}}\Big)\Big(\prod_{i<j}^3  Z_{ij}^{-\Delta_{ij}}\Big)\\\equiv &P_1\Big(\prod_{i<j}^3  Z_{ij}^{-\Delta_{ij}}\Big)
\ea\ee
and
\be\ba \label{p2}
&G\exp\Big(\sum_{i<j} A_{ij}\frac{\theta_{ij}\bar{\th}_{ij}}{z_{ij}}\Big)\\
=&\sum_{j,k,j\neq k}\Big(2\frac{\th_{0k}\bar{\th}_{0k}}{z_{0k}}A_{jk}\frac{\theta_{jk}\bar{\th}_{jk}}{z_{jk}^2}+\frac{\th_{0k}}{Z_{0k}} A_{kj} \frac{\bar{\th}_{kj}}{Z_{kj}} -\frac{\bar{\th}_{0k}}{Z_{0k}} A_{jk} \frac{ \th_{jk}}{Z_{jk}} \Big)\exp\Big(\sum_{i<j} A_{ij}\frac{\theta_{ij}\bar{\th}_{ij}}{z_{ij}}\Big)\\
\equiv&P_2 \exp\Big(\sum_{i<j} A_{ij}\frac{\theta_{ij}\bar{\th}_{ij}}{z_{ij}}\Big).
\ea\ee
Thus we obtain $
\vev{J\Phi_1\Phi_2\Phi_3}=(F+P_1+P_2)\vev{\Phi_1\Phi_2\Phi_3}$,
and it follows that
\be\ba\label{JJ223}
\vev{J\bar{J}\Phi_1\Phi_2\Phi_3}=&(G+F)(\tilde{G}+\tilde{F})\vev{\Phi_1\Phi_2\Phi_3}\\
=&(P_1+P_2+F)(\tilde{P}_1+\tilde{P}_2+\tilde{F})\vev{\Phi_1\Phi_2\Phi_3}+[G(\tilde{P}_1+\tilde{P}_2+\tilde{F})]\vev{\Phi_1\Phi_2\Phi_3},
\ea\ee
where the last term should be dropped as discussed in previous sections. Substituting the expression of $P_1,P_2,F$ into eq.(\ref{JJ223}), we have
\be\ba \label{3ptint}
&\frac{1}{\vev{\Phi_1\Phi_2\Phi_3}}\int d^2zd\th d\bar{\th}d\tt d\btt \vev{J\bar{J}\Phi_1\Phi_2\Phi_3}\\
=&\int d^2z \Big[\sum_{i,k,i\neq k}\Big(-\frac{2}{z_{0k}}\frac{\Delta_{ik}}{Z_{ik}}+\frac{1}{z^2_{0k}}\frac{(\th_k\bar{\th}_{i}+\bar{\th}_k\th_i)\Delta_{ik}}{Z_{ik}}-2\frac{1}{z_{0k}}A_{ik}\frac{\theta_{ik}\bar{\th}_{ik}}{z_{ik}^2}-\frac{\th_{k}}{z^2_{0k}} A_{ki} \frac{\bar{\th}_{ki}}{Z_{ki}}\\
&-\frac{\bar{\th}_{k}}{z^2_{0k}} A_{ik} \frac{ \th_{ik}}{Z_{ik}} \Big) +\sum_i\Big(-2\Delta_i\frac{1}{z^2_{0i}}+2\frac{Q_i\bar{\th}_i\th_i}{z_{0i}^3}\Big)\Big]\\
&\times\Big[\sum_{j,k,j\neq k}\Big(-\frac{2}{\bar{z}_{0k}}\frac{\bar{\Delta}_{jk}}{\tilde{Z}_{jk}}+\frac{1}{\bar{z}^2_{0k}}\frac{(\tt_k\bar{\tt}_{j}+\bar{\tt}_k\tt_j)\bar{\Delta}_{jk}}{\tilde{Z}_{jk}}-2\frac{1}{\bar{z}_{0k}}\bar{A}_{jk}\frac{\tt_{jk}\bar{\tt}_{jk}}{\bar{z}_{jk}^2}-\frac{\tt_{k}}{\bar{z}^2_{0k}} \bar{A}_{kj} \frac{\bar{\tt}_{kj}}{\bar{Z}_{kj}} \\
&-\frac{\bar{\tt}_{k}}{\bar{z}^2_{0k}} \bar{A}_{jk} \frac{ \tt_{jk}}{\bar{Z}_{jk}} \Big)+\sum_j\Big(-2\bar{\Delta}_j\frac{1}{\bar{z}^2_{0j}}+2\frac{\bar{Q}_j\btt_j\tt_j}{\bar{z}_{0j}^3}\Big)\Big].
\ea\ee
Using $\sum_{i,i\neq k}\Delta_{ik}=2\Delta_k$, the first and second line of the integrand can be expressed as
\be\ba \label{sort}
&\Big[\sum_{i,k,i\neq k}\Big(-\frac{2}{z_{0i}}\frac{\Delta_{ki}}{Z_{ki}}+\frac{1}{z^2_{0i}}\frac{(\th_i\bar{\th}_{k}+\bar{\th}_i\th_k)\Delta_{ki}}{Z_{ki}}-2\frac{1}{z_{0i}}A_{ki}\frac{\theta_{ki}\bar{\th}_{ki}}{z_{ki}^2}-\frac{\th_{i}}{z^2_{0i}} A_{ik}\Big(\frac{\bar{\th}_{ik}}{z_{ik}}-\frac{\th_{ik}\bar{\th}_i\bar{\th}_k}{z^2_{ik}}\Big)\\
&-\frac{\bar{\th}_{i}}{z^2_{0i}} A_{ki}\Big(\frac{ \th_{ki}}{z_{ki}}-\frac{\th_k\th_i\bar{\th}_{ki}}{z^2_{ik}}\Big)\Big)+\sum_i\Big( -2\Delta_i\frac{1}{z^2_{0i}}+2\frac{Q_i\bar{\th}_i\th_i}{z_{0i}^3}\Big)\Big]\\
=& \sum_{i,k,i\neq k}\Big(-\frac{1}{z_{0i}}\Big(\frac{2\Delta_{ki}}{Z_{ki}}+\frac{2\th_{ki}\bar{\th}_{ki}A_{ki}}{z^2_{ki}}\Big)-\frac{1}{z^2_{0i}}\Big(\frac{z_{ki}\Delta_{ki}}{Z_{ki}}+A_{ik} \frac{\th_i\bar{\th}_{ik}-\bar{\th}_i\th_{ik}}{z_{ik}}\Big)\Big)+\sum_i\frac{2Q_i\bar{\th}_i\th_i}{z^3_{0i}}.
\ea\ee
\subsection{$\mathcal{N}=(2,2)$: $n$-point case}\label{details6}
 Let us first consider only holomorphic component $J(Z)$ inserted
\be\label{Jn2}
\vev{J\Phi_1...\Phi_n}=(G+F)\vev{\Phi_1...\Phi_n},
\ee
where we will encounter new crossing terms $G \tilde{R}_{ijk}$, $G \tilde{Z}_{ijkl}$ in addition to these appeared in eq.(\ref{cross2}). By using eq.(\ref{delde}) it can be checked that they will vanish, i.e.
\be \ba
\int d\th d\bar{\th}G \tilde{R}_{ijk}=0,~~\int d\th d\bar{\th}G \tilde{Z}_{ijkl}=0.
\ea\ee
Thus we will not consider crossing terms in eq.(\ref{Jn2}), then
\be \ba
&(G+F)\vev{\Phi_1...\Phi_n}= (P_1+P_2+F)\vev{\Phi_1...\Phi_n}\\
=&(P_1+P_2+F)\vev{\Phi_1...\Phi_n}+ Q\vev{\Phi_1...\Phi_n}\frac{1}{f},
\ea\ee
where $P_1,P_2$ be of the same form as defined in eq.(\ref{p1}) and eq.(\ref{p2}). Here $Q$ equals $Gf$, which is
\be \ba\label{Qdef}
Q=&\Big(\sum_{R_{ijk}}(GR_{ijk})\frac{\p f}{\p R_{ijk}}+\sum_{Z_{ijkl}}(GZ_{ijkl})\frac{\p f}{\p Z_{ijkl}}\Big)\\
=&\sum_n \Big(\frac{2\th_{0n}\bar{\th}_{0n}}{z_{0n}}\Big((\p_{z_n}R)\frac{\p f}{\p R} +(\p_{z_n}Z)\frac{\p f}{\p Z} \Big)+\frac{\th_{0n}}{Z_{0n}}\Big((\p_{\th_n}R)\frac{\p f}{\p R} +(\p_{\th_n}Z)\frac{\p f}{\p Z} \Big)\\
&-\frac{\bar{\th}_{0n}}{Z_{0n}}\Big((\p_{\bar{\th}_n}R)\frac{\p f}{\p R} +(\p_{\bar{\th}_n}Z)\frac{\p f}{\p Z} \Big)\Big)\\
\equiv&\sum_n \Big(\frac{2\th_{0n}\bar{\th}_{0n}}{z_{0n}}\p_{z_n}^R f+\frac{\th_{0n}}{Z_{0n}}\p_{\th_n}^R f-\frac{\bar{\th}_{0n}}{Z_{0n}}\p_{\bar{\th}_{n}}^R f\Big),
\ea\ee
where for simplicity we have abbreviated $R_{ijk}$ as $R$, $Z_{ijkl}$
as $Z$ and suppressed the summation $\sum_{R_{ijk}},\sum_{Z_{ijkl}}$. Note in the first step in eq.(\ref{Qdef}) we omit the terms vanishing after integration over $\th,\bar{\th}$.  Following the same way we introduce $\tilde{Q}$ as
\be \ba
\tilde{Q}&\equiv \sum_n \Big(\frac{2\tt_{0n}\bar{\tt}_{0n}}{\bar{z}_{0n}}\p_{\bar{z}_n}^L f+\frac{\tt_{0n}}{\tilde{Z}_{0n}}\p_{\tt_n}^L f-\frac{\bar{\tt}_{0n}}{\tilde{Z}_{0n}} \p_{\bar{\tt}_{n}}^L f\Big).
\ea\ee
Next consider $\vev{J\bar{J}\Phi_1...\Phi_n}$, which is
\be \ba
&(G+F)(\tilde{G}+\tilde{F})\vev{\Phi_1...\Phi_n}\\=&(F+P)(\tilde{F}+\tilde{F})\vev{\Phi_1...\Phi_n}+Q(\tilde{F}+\tilde{P})\vev{\Phi_1...\Phi_n}/f+(F+P)\tilde{Q}\vev{\Phi_1...\Phi_n}/f\\
&+(G\tilde{Q})\vev{\Phi_1...\Phi_n}/f+[G(\tilde{P}+\tilde{F})]\vev{\Phi_1...\Phi_n},
\ea\ee
where the last term should be dropped as discussed in previous sections. And the term $(G\tilde{Q})\vev{\Phi_1...\Phi_n}/f$   is very similar to the (1,1) case as discussed in eq.(\ref{GQ}), which is not a crossing term. Actually,
\be\ba \label{GtQ}
 G\tilde{Q}=& \sum_{i,n} \Big(\frac{2\th_{0i}\bar{\th}_{0i}}{z_{0i}}\p^R_{z_i}+\frac{\th_{0i}}{Z_{0i}}\p_{\th_i}^R-\frac{\bar{\th}_{0i}}{Z_{0i}}\p_{\bar{\th}_i}^R\Big)\Big(\frac{2\tt_{0n}\bar{\tt}_{0n}}{\bar{z}_{0n}}\p_{\bar{z}_n}^L +\frac{\tt_{0n}}{\tilde{Z}_{0n}}\p_{\tt_n}^L -\frac{\bar{\tt}_{0n}}{\tilde{Z}_{0n}} \p_{\bar{\tt}_{n}}^L \Big) f,
\ea\ee
thus
\be\ba
&\int  d\th d\bar{\th}d\tt d\bar{\tt} G\tilde{Q}= \sum_{i,n}\Big(
\frac{2}{z_{0i}}\p^R_{z_i}+\frac{\th_{i}}{z^2_{0i}}\p_{\th_i}^R+\frac{\bar{\th}_{i}}{z^2_{0i}}\p_{\bar{\th}_i}^R\Big)\Big(\frac{2 }{\bar{z}_{0n}}\p_{\bar{z}_n}^L +\frac{\tt_{n}}{\tilde{z}^2_{0n}}\p_{\tt_n}^L+\frac{\bar{\tt}_{n}}{\tilde{z}^2_{0n}} \p_{\bar{\tt}_{n}}^L \Big)f.
\ea\ee
Gathering all the results together, we then have
{\small{
\be\ba \label{preint}
&\frac{1}{\vev{\Phi_1...\Phi_n}}\int d^2zd\th d\bar{\th}d\tt d\btt \vev{J\bar{J}\Phi_1...\Phi_n}\\
=&\int d^2z \Big[\sum_{i,k,i\neq k}\Big(-\frac{2}{z_{0k}}\frac{\Delta_{ik}}{Z_{ik}}+\frac{1}{z^2_{0k}}\frac{(\th_k\bar{\th}_{i}+\bar{\th}_k\th_i)\Delta_{ik}}{Z_{ik}}-2\frac{1}{z_{0k}}A_{jk}\frac{\theta_{jk}\bar{\th}_{jk}}{z_{jk}^2}-\frac{\th_{k}}{z^2_{0k}} A_{kj} \frac{\bar{\th}_{kj}}{Z_{kj}}
\\&-\frac{\bar{\th}_{k}}{z^2_{0k}} A_{jk} \frac{ \th_{jk}}{Z_{jk}}\Big) +\sum_i\Big(-2\Delta_i\frac{1}{z^2_{0i}}+2\frac{Q_i\bar{\th}_i\th_i}{z_{0i}^3}\Big)\Big]\\
&\times \Big[\sum_{i,k,i\neq k}\Big(-\frac{2}{\bar{z}_{0k}}\frac{\bar{\Delta}_{ik}}{\tilde{Z}_{ik}}+\frac{1}{\bar{z}^2_{0k}}\frac{(\tt_k\bar{\tt}_{i}+\bar{\tt}_k\tt_i)\bar{\Delta}_{ik}}{\tilde{Z}_{ik}}-2\frac{1}{\bar{z}_{0k}}\bar{A}_{jk}\frac{\tt_{jk}\bar{\tt}_{jk}}{\bar{z}_{jk}^2}-\frac{\tt_{k}}{\bar{z}^2_{0k}} \bar{A}_{kj} \frac{\bar{\tt}_{kj}}{\bar{Z}_{kj}}
\\&-\frac{\bar{\tt}_{k}}{\bar{z}^2_{0k}} \bar{A}_{jk }\frac{ \tt_{jk}}{\bar{Z}_{jk}} \Big) +\sum_i\Big(-2\bar{\Delta}_i\frac{1}{\bar{z}^2_{0i}}+2\frac{\bar{Q}_i\btt_i\tt_i}{\bar{z}_{0i}^3}\Big)\Big]+\sum_n \Big(\frac{-2}{z_{0n}}\p_{z_n}^R f-\frac{\th_{n}}{z^2_{0n}}\p_{\th_n}^R f-\frac{\bar{\th}_{n}}{z^2_{0n}}\p_{\bar{\th}_{n}}^R f\Big)\frac{1}{f}\\
&\times \Big[\sum_{i,k,i\neq k}\Big(-\frac{2}{\bar{z}_{0k}}\frac{\bar{\Delta}_{ik}}{\tilde{Z}_{ik}}+\frac{1}{\bar{z}^2_{0k}}\frac{(\tt_k\bar{\tt}_{i}+\bar{\tt}_k\tt_i)\bar{\Delta}_{ik}}{\tilde{Z}_{ik}}-2\frac{1}{\bar{z}_{0k}}\bar{A}_{jk}\frac{\tt_{jk}\bar{\tt}_{jk}}{\bar{z}_{jk}^2}-\frac{\tt_{k}}{\bar{z}^2_{0k}} \bar{A}_{kj} \frac{\bar{\tt}_{kj}}{\bar{Z}_{kj}} \\
&-\frac{\bar{\tt}_{k}}{\bar{z}^2_{0k}} \bar{A}_{jk} \frac{ \tt_{jk}}{\bar{Z}_{jk}} \Big) +\sum_i\Big(-2\bar{\Delta}_i\frac{1}{\bar{z}^2_{0i}}+2\frac{\bar{Q}_i\btt_i\tt_i}{\bar{z}_{0i}^3}\Big)\Big]\\
&+
\Big[\sum_{i,k,i\neq k}\Big(-\frac{2}{z_{0k}}\frac{\Delta_{ik}}{Z_{ik}}+\frac{1}{z^2_{0k}}\frac{(\th_k\bar{\th}_{i}+\bar{\th}_k\th_i)\Delta_{ik}}{Z_{ik}}-2\frac{1}{z_{0k}}A_{jk}\frac{\theta_{jk}\bar{\th}_{jk}}{z_{jk}^2}-\frac{\th_{k}}{z^2_{0k}} A_{kj} \frac{\bar{\th}_{kj}}{Z_{kj}} \\
&-\frac{\bar{\th}_{k}}{z^2_{0k}} A_{jk} \frac{ \th_{jk}}{Z_{jk}} \Big) +\sum_i\Big(-2\Delta_i\frac{1}{z^2_{0i}}+2\frac{Q_i\bar{\th}_i\th_i}{z_{0i}^3}\Big)\Big]\times  \sum_n \Big(\frac{-2}{\bar{z}_{0n}}\p_{\bar{z}_n}^L f-\frac{\tt_{n}}{\bar{z}^2_{0n}}\p_{\tt_n}^L f-\frac{\bar{\tt}_{n}}{\bar{z}^2_{0n}} \p_{\bar{\tt}_{n}}^L f\Big)\frac{1}{f}\\
&+\int d^2z \Big(\sum_{i,n}\Big(
\frac{2}{z_{0i}}\p^R_{z_i}+\frac{\th_{i}}{z^2_{0i}}\p_{\th_i}^R+\frac{\bar{\th}_{i}}{z^2_{0i}}\p_{\bar{\th}_i}^R\Big)\Big(\frac{2 }{\bar{z}_{0n}}\p_{\bar{z}_n}^L +\frac{\tt_{n}}{\tilde{z}^2_{0n}}\p_{\tt_n}^L+\frac{\bar{\tt}_{n}}{\tilde{z}^2_{0n}} \p_{\bar{\tt}_{n}}^L \Big)f\Big)\frac{1}{f}\\
\ea\ee}}
Note that the first term of the integrand has the same form as 3-pt correlators in eq.(\ref{3ptint}) except for the summation here runs from 1 to $n$ instead of 3 in eq.(\ref{3ptint}).
\section{Integrals in 2-point correlators }\label{appb}
There are nine terms in eq.(\ref{intJJ}), the first one have been considered in eq.(\ref{T11}). Below by using the integrals in section \ref{DR} we list the remaining eight terms in the integral eq.(\ref{intJJ}).
The second term
\be
\ba
T_{22}\equiv&\int d^2z d\theta d\bar{\theta} \frac{\Delta\theta_{12}}{Z_{12}}\Big( \frac{1}{Z_{01}}+\frac{1}{Z_{02}}\Big)\frac{\Delta\bar{\theta}_{12}}{\bar{Z}_{12}}\Big( \frac{1}{\bar{Z}_{01}}+\frac{1}{\bar{Z}_{02}}\Big) \\
=&\frac{\Delta^2\theta_{12} \bar{\theta}_{12}}{Z_{12}\bar{Z}_{12}}\int d^2z \int d\theta \Big( \frac{1}{Z_{01}}+\frac{1}{Z_{02}}\Big) \int d\bar{\theta}\Big( \frac{1}{\bar{Z}_{01}}+\frac{1}{\bar{Z}_{02}}\Big)\\
=&\frac{\Delta^2\theta_{12} \bar{\theta}_{12}}{Z_{12}\bar{Z}_{12}}\int d^2z  \Big( \frac{\theta_1}{z_{01}^2}+\frac{\theta_2}{z_{02}^2}\Big)   \Big( \frac{\bar{\theta}_1}{\bar{z}_{01}^2}+\frac{\bar{\theta}_2}{\bar{z}_{02}^2}\Big)\\
=&-\frac{\Delta^2\theta_{1}\theta_2\bar{\theta}_1\bar{\theta}_2 }{Z_{12}\bar{Z}_{12}}(2\CI_{22}(z_1,\bar{z}_1)+\CI_{22}(z_1,\bar{z}_2)+\CI_{22}(z_2,\bar{z}_1))=0.
\ea
\ee
The third term
\be
\ba
T_{33}\equiv&\Delta^2\int d^2z d\theta d\bar{\theta}  \Big( \frac{\theta_{01}}{Z_{01}^2}+\frac{\theta_{02}}{Z_{02}^2}\Big) \Big( \frac{\bar{\theta}_{01}}{\bar{Z}_{01}^2}+\frac{\bar{\theta}_{02}}{\bar{Z}_{02}^2}\Big) \\
=&-\Delta^2\int d^2z   \Big( \frac{1}{z_{01}^2}+\frac{1}{z_{02}^2}\Big) \Big( \frac{ 1}{\bar{z}_{01}^2}+\frac{1}{\bar{z}_{02}^2}\Big) \\
=&-\Delta^2 (2\CI_{22}(z_1,\bar{z}_1)+\CI_{22}(z_1,\bar{z}_2)+\CI_{22}(z_2,\bar{z}_1))=0.
\ea
\ee
The fourth term
\be
\ba
T_{12}\equiv &\int d^2z d\theta d\bar{\theta}\frac{2\Delta^2}{Z_{12}}\Big( \frac{\theta_{01}}{Z_{01}}- \frac{\theta_{02}}{Z_{02}}\Big)\frac{\bar{\theta}_{12}}{\bar{Z}_{12}}\Big( \frac{1}{\bar{Z}_{01}}+\frac{1}{\bar{Z}_{02}}\Big)\\
=&\frac{2\Delta^2 \bar{\theta}_{1}\bar{\theta}_2  }{Z_{12}\bar{Z}_{12}}\int d^2z\Big(\frac{1}{z_{01}}- \frac{1}{z_{02}}\Big)\Big(\frac{1}{\bar{z}_{01}^2}+ \frac{1}{\bar{z}_{02}^2}\Big)\\
=&\frac{2\Delta^2 \bar{\theta}_{1}\bar{\theta}_2   }{Z_{12}\bar{Z}_{12}} (\CI_{12}(z_1,\bar{z}_1)+\CI_{12}(z_1,\bar{z}_2)-\CI_{12}(z_2,\bar{z}_1)-\CI_{12}(z_2,\bar{z}_2))\\
=&\frac{2\Delta^2 \bar{\theta}_{1}\bar{\theta}_2   }{Z_{12}\bar{Z}_{12}} ( \CI_{12}(z_1,\bar{z}_2)-\CI_{12}(z_2,\bar{z}_1) )=\frac{2\Delta^2 \bar{\theta}_{1}\bar{\theta}_2   }{Z_{12}\bar{Z}_{12}}\frac{2\pi}{\bar{z}_{12}}.
\ea
\ee
The fifth term
\be
T_{21}\equiv \frac{2\Delta^2 \theta_{1}\theta_2}{Z_{12}\bar{Z}_{12}} ( \CI_{12}(\bar{z}_1,z_2)-\CI_{12}(\bar{z}_2,z_1) )=\frac{2\Delta^2 \theta_{1}\theta_2}{Z_{12}\bar{Z}_{12}} \frac{2\pi}{z_{12}}.
\ee
The sixth term
\be
\ba
T_{13}\equiv &-\int d^2z d\theta d\bar{\theta}\frac{2\Delta^2}{Z_{12}}\Big( \frac{\theta_{01}}{Z_{01}}- \frac{\theta_{02}}{Z_{02}}\Big) \Big( \frac{\bar{\theta}_{01}}{\bar{Z}_{01}^2}+\frac{\bar{\theta}_{02}}{\bar{Z}_{02}^2}\Big)\\
=&\frac{2\Delta^2    }{Z_{12} }\int d^2z\Big(\frac{1}{z_{01}}- \frac{1}{z_{02}}\Big)\Big(\frac{1}{\bar{z}_{01}^2}+ \frac{1}{\bar{z}_{02}^2}\Big)\\
=&\frac{2\Delta^2    }{Z_{12} }( \CI_{12}(z_1,\bar{z}_2)-\CI_{12}(z_2,\bar{z}_1) )=\frac{2\Delta^2    }{Z_{12} }\frac{2\pi}{\bar{z}_{12}}.
\ea
\ee
The seventh term
\be
\ba
T_{31}\equiv \frac{2\Delta^2    }{\bar{Z}_{12} }( \CI_{12}(\bar{z}_1,z_2)-\CI_{12}(\bar{z}_2,z_1) )=\frac{2\Delta^2    }{\bar{Z}_{12} } \frac{2\pi}{z_{12}}.
\ea
\ee
The eighth term
\be
\ba
T_{23}\equiv &-\int d^2z d\theta d\bar{\theta}\frac{\Delta^2\theta_{12}}{Z_{12}}\Big( \frac{1}{Z_{01}}+ \frac{1}{Z_{02}}\Big) \Big( \frac{\bar{\theta}_{01}}{\bar{Z}_{01}^2}+\frac{\bar{\theta}_{02}}{\bar{Z}_{02}^2}\Big)\\
=&-\frac{\Delta^2  \theta_1\theta_2  }{Z_{12} }\int d^2z\Big(\frac{1}{z_{01}^2}+ \frac{1}{z^2_{02}}\Big)\Big(\frac{1}{\bar{z}_{01}^2}+ \frac{1}{\bar{z}_{02}^2}\Big)\\
=&-\frac{\Delta^2  \theta_1\theta_2  }{Z_{12} }(2\CI_{22}(z_1,\bar{z}_1)+\CI_{22}(z_1,\bar{z}_2)+\CI_{22}(z_2,\bar{z}_1))=0.
\ea
\ee
The ninth term
\be
\ba
T_{32}\equiv&  -\frac{\Delta^2  \bar{\theta}_1\bar{\theta}_2  }{\bar{Z}_{12} }(2\CI_{22}(z_1,\bar{z}_1)+\CI_{22}(z_1,\bar{z}_2)+\CI_{22}(z_2,\bar{z}_1))=0.
\ea
\ee
Finally, the total contribution from the eight terms is
\be\ba
T_{12}+T_{21}+T_{13}+T_{23}=\frac{8\pi\Delta^2}{Z_{12}\bar{Z}_{12}}.
\ea\ee

\end{document}